\documentclass{emulateapj}

\usepackage{amssymb}
\usepackage{amsmath}
\usepackage{epstopdf}
\usepackage{natbib}
\usepackage[colorlinks, citecolor=blue]{hyperref}
\usepackage{hyperref}
\usepackage[all]{hypcap}

\begin{document}

\title{A Monte Carlo Approach to Magnetar-powered Transients: I.
Hydrogen-deficient Superluminous Supernovae}
\author{Liang-Duan Liu\altaffilmark{1,2}, Shan-Qin Wang\altaffilmark{1,2,3},
Ling-Jun Wang\altaffilmark{4}, Zi-Gao Dai\altaffilmark{1,2}, Hai Yu%
\altaffilmark{1,2}, Zong-Kai Peng\altaffilmark{1,2}}

\begin{abstract}
In this paper we collect 19 hydrogen-deficient superluminous supernovae
(SLSNe) and fit their light curves, temperature evolution, and velocity
evolution based on the magnetar-powered model. To obtain the best-fitting
parameters, we incorporate the Markov Chain Monte Carlo approach. We get
rather good fits for 7 events ($\chi^2$/d.o.f = $0.24-0.96$) and good fits
for other 7 events ($\chi^2$/d.o.f = $1.37-3.13$). We find that the initial
periods ($P_{0}$) and magnetic strength ($B_{p}$) of the magnetars supposed
to power these SLSNe are in the range of $\sim 1.2-8.3$ ms and $\sim
(0.2-8.8)\times 10^{14}$ G, respectively; the inferred masses of the ejecta
of these SLSNe are between 1 and 27.6 $M_{\odot }$, and the values of the
gamma-ray opacity $\kappa_{\gamma }$ are between 0.01 and 0.82 cm$^2$ ~g$%
^{-1}$. We also calculate the fraction of the initial rotational energy of
the magnetars harbored in the centers of the remnants of these SLSNe that is
converted to the kinetic energy of the ejecta and find that the fraction is $%
\sim 19-97$\% for different values of $P_{0}$ and $B_{p}$, indicating that
the acceleration effect cannot be neglected. Moreover, we find that the
initial kinetic energies of most of these SLSNe are so small ($\lesssim
2\times 10^{51}$ erg) that they can be easily explained by the
neutrino-driven mechanism. These results can help clarify some important
issues related to the energy-source mechanisms and explosion mechanisms and
reveal the nature of SLSNe.
\end{abstract}


\affil{\altaffilmark{1}School of Astronomy and Space Science, Nanjing
University, Nanjing 210093, China; dzg@nju.edu.cn}
\affil{\altaffilmark{2}Key Laboratory of Modern Astronomy and Astrophysics (Nanjing
University), Ministry of Education, China}
\affil{\altaffilmark{3}Department
of Astronomy, University of California, Berkeley, CA 94720-3411, USA}
\affil{\altaffilmark{4}Key Laboratory of Space Astronomy and Technology,
National Astronomical Observatories, Chinese Academy of Sciences, Beijing
100012, China}

\section{Introduction}

\label{sec:Intro}

Superluminous supernovae \citep[SLSNe;][]{Qui2011,Gal2012}, whose peak
luminosities are $\gtrsim 10^{44}$ erg s$^{-1}$, have been discovered and
studied in the last decade. According to their optical spectra near maximum
light, SLSNe can be divided into types I (hydrogen-deficient) and II
(hydrogen-rich; the detailed classification schemes of SNe based on their
spectra were summarized by \citealt{Fil1997} and \citealt{Gal2016}). To
date, all SLSNe-I are type Ic SNe whose spectral features resemble that of
the normal-luminosity SNe Ic \citep{Pas2010,Gal2012,Inse2013,Nich2016b}.
While most SLSNe-II have narrow H$\alpha$ emission lines in their optical
spectra and can be regarded as the high-luminosity versions of normal SNe
IIn, a few SLSNe II do not have H$\alpha$ emission lines and their spectra
resemble those of SNe IIL \citep{Gez2009,Mil2009,Ins2016}.

The problems of the origin of the energy sources and explosion mechanisms of
SLSNe have not yet been completely solved. Currently, the most prevailing
energy-source models explaining SLSNe are magnetar-powered model %
\citep[e.g.,][]{Kas2010,Woos2010,Des2012,Inse2013,Chen2015,Wang2015a,Wang2016a,Dai2016,Wang2016c}
\footnote{%
It has long been recognized that the dipole radiation from the nascent
neutron stars can enhance the luminosities of the normal SNe %
\citep{Ost1971,Mae2007}.}, the ejecta-circumstellar medium (CSM) interaction
model \citep{Che1982,Che1994,Chu1994,Chu2009,Che2011,Cha2012,Gin2012}, and
the pair instability supernova (PISN) model \citep{Bar1967,Rak1967} which is
essentially $^{56}$Ni-powered \citep{Col1969,Col1980,Arn1982} but requires a
huge amount ($\gtrsim 5M_{\odot }$) of $^{56}$Ni. Some SLSNe show
double-peaked light curves \citep{Nich2015a,NS2015,Smit2016,Vre2017} and
their early-time excess emission might be due to the cooling emission from
shock-heated envelopes of the progenitors while the main peaks can be
explained by the magnetar-powered model or interaction model.

Determining the energy-source models for SLSNe is rather tricky. For
example, \citet{Gal2009} proposed that SN 2007bi, whose light-curve decline
rate is approximately equal to the decay rate of $^{56}$Co, is a PISN, but %
\citet{Des2012} argued that it might not be a PISN since its spectrum is
inconsistent with the spectrum produced by the PISN model; \citet{Nich2013}
demonstrated that another slow declining SLSN PTF12dam, whose post-peak
light curve mimics that reproduced by the PISN model, is not yet a PISN
since the rising part of its light curve cannot be explained by the PISN
model.

Except for the two high-redshift SLSNe (SN 2213-1745 and SN 1000+0216) that
are believed to be PISNe \citep{Cooke2012} and SN 2007bi whose explosion
mechanism is still in debate,\footnote{\citet{Inse2017} assembled four
low-redshift slow-evolving SLSNe I (SN 2007bi, PTF12dam, SN 2015bn, and
LSQ14an) and found the declines of their light curves at $t-t_{\text{peak}}
\gtrsim 150$ days are faster than $^{56}$Co and further steepen after 300
days, which indicates that these four SLSNe I cannot be explained by the
PISN model since the required ejecta are so massive that the decline rates
of the light curves must be consistent with that of $^{56}$Co decay at $t-t_{%
\text{peak}}\gtrsim 500$ days. A similar analysis cannot be performed for
high-redshift analogues since the lower quality late-time data prevent
further investigation, and therefore the possibility that high-redshift
slow-evolving SLSNe are PISNe cannot be excluded.} all SLSNe discovered
cannot be explained by the PISN model since the ratio of required masses of $%
^{56}$Ni to inferred ejecta masses are too large and/or the theoretical
light curves did not fit the observational data %
\citep{Qui2011,Inse2013,Nich2013,Nich2014}. This fact in turn indicates that
most SLSNe observed might be core-collapse SNe (CCSNe) since the peak
luminosities of type Ia SNe cannot reach $10^{44}$ erg s$^{-1}$. \footnote{%
Type Ia SNe magnified by the gravitational lensing effect, e.g., PS1-10afx %
\citep{Qui2013,Qui2014}, are not genuine SLSNe.}

SLSNe that cannot be explained by PISN model can be explained by the
magnetar-powered model \citep{Inse2013,Nich2013,Nich2014} or the ejecta-CSM
interaction model \citep[e.g.][]{Smi2007,Mor2013a,Nich2014}. The
magnetar-powered model proposes that the nascent neutron stars with initial
rotational periods of a few milliseconds and magnetic strength of $%
10^{13}-10^{15}$ G inject their rotational energy to the ejecta of SNe,
while the ejecta-CSM interaction model supposes that the interaction between
the SN ejecta and the CSM surrounding the progenitors of the SNe can release
a huge amount of energy and heat the SN ejecta. These two processes can
significantly enhance the luminosities of the SNe and lead them to be SLSNe.

In this paper we focus on the SLSNe that were demonstrated not to be PISNe.
While the ejecta-CSM model cannot be excluded in explaining these SLSNe Ic,
we do not adopt it due to the absence of the interaction signatures (e.g.,
narrow H$_{\alpha }$ emission lines) that are indicative of strong
interaction between the ejecta and CSM. Studies have shown that the
magnetar-powered model can well reproduce the light curves of these SLSNe,
and indicated that the magnetar-powered model has special advantages since
it does not need ad hoc assumptions about the pre-supernova mass-loss
history. Therefore, the magmetar model is preferred in modeling SLSNe Ic.

Most previous studies using the magnetar-powered model focused on the
light-curve fitting. Some group fitted the light curves, temperature
evolution and the evolution of photospheric radii of SLSNe. However, the
photospheric radius of an SLSN cannot be measured directly, and it is
derived from the luminosity and temperature ($L=4\pi \sigma T^4R^2$, where $%
L $ is the luminosity, $\sigma$ is the Stefan-Boltzman constant, $T$ is the
temperature, and $R$ is the photospheric radius).

In this paper, we collect a sample of 19 type I SLSNe and use the
manetar-powered model proposed by \citet{Wang2016a} to fit their light
curves, temperature evolution, and velocity evolution. To obtain the
best-fitting parameters, we use the Markov Chain Monte Carlo (MCMC) code
developed by \citet{Wang2016b}, who employed this approach (manetar-powered
model + MCMC) to fit the light curves of SNe 1998bw and 2002ap.

This paper is organized as follows. In Section \ref{sec:mod}, we present our
sample and describe the magnetar-powered model adopted. In Section \ref%
{sec:res}, we use the magmetar-powered model and the MCMC code to get the
best-fitting parameters. We discuss implications of our results in Section %
\ref{sec:dis} and give some conclusions in Section \ref{sec:con}. In our
another paper \citep{Wang2017}, we systematically study the whole sample of
broad-lined type Ic supernovae not associated with gamma-ray bursts. We find
that the magnetar-powered model can also well account for both the light
curves and velocity evolution of these SNe.

\section{Sampling and Modeling}

\label{sec:mod}

We collect 19 type I (Ic) SLSNe in the literature, see Table \ref{tbl:sample}%
. All of these SLSNe have light curves and temperature data, and most of
them have velocity data. These SLSNe are selected because the observational
errors were provided by the original papers so that we can run the MCMC code
against them.

The most prevailing magnetar model are the model proposed by \citet{Cha2012}
and \citet{Inse2013} the model revised by \citet{Wang2015a} and %
\citet{Chen2015} by incorporating the hard emission leakage effect. These
semi-analytical magnetar-powered models used to fit most type I SLSNe
neglect the photospheric recession effect and acceleration of the SN ejecta
caused by the magnetar wind. \citet{Kas2010} and \citet{Woos2010} proposed
the magnetar-powered model for SLSNe and demonstrated the acceleration
effect is rather notable but they do not take into account the leakage
effect. \citet{Wang2016a} proposed a revised magnetar model that took into
account all these effects (leakage effect, photospheric recession effect and
acceleration effect). So we use this revised magnetar-powered model to fit
the observed data.

In this revised magnetar-powered model, the photospheric velocities of SLSNe
cannot be fixed to be their scale velocities $v_{\text{sc}}$ and their
evolution must be fitted. Moreover, the scale velocity itself is a varying
quantity and not directly measurable. Instead, the initial scale velocity $%
v_{\text{sc0}}$ is a free parameter. If the neutrinos emitted from
proto-neutron stars provide the initial kinetic energy (KE) and initial
velocity, then the values of $v_{\text{sc0}}$ must have a lower limit to
ensure that the initial KE of SNe can be larger than the lower limit of the
initial KE provided by neutrinos
\citep[$\sim10^{50}$ erg,][and references therein]{Jan2016}.

Other parameters in this model include the mass of ejecta $M_{\text{ej}}$,
grey optical opacity $\kappa $, magnetic strength of magnetar $B_{p}$,
initial rotational period of the magnetar $P_{0}$, and the gamma-ray opacity
$\kappa _{\gamma }$ to magnetar photons.\footnote{%
In order to quantitatively describe the leakage effect associated with gamma
and X-ray photons emitted from the magnetar, \citet{Wang2015a} incorporated
the trapping factor ($1-e^{-At^{-2}}$) into the original magnetar-powered
model, $A=3\kappa _{\gamma }M_{\text{ej}}/4\pi v^{2}$, $\kappa _{\gamma }$
is the effective gamma ray opacity.} Here the subscript \textquotedblleft $p$%
" in $B_{p}$ means the surface dipole magnetic field of the magnetar %
\citep{Shapiro83}. The optical opacity $\kappa $ is somewhat uncertain and
we adopt the value 0.1 cm$^{2}$ g$^{-1}$ throughout this paper. 
Because the explosion time of an SLSN is not an observable quantity, a free
parameter $T_{\text{start}}$ is included to refer to the
theoretical explosion time relative to the zero epoch given in the paper
from which the observed data have been taken. In the papers providing the data,
sometimes the zero epochs refer to the peak times, and otherwise the zero epochs refer to
the inferred explosion times.

While the PISN model (pure $^{56}$Ni-powered model) has been excluded by
previous studies for most type I SLSNe, a moderate amount of $^{56}$Ni
synthesized by the these SLSNe cannot be completely neglected. However, the
masses of $^{56}$Ni synthesized by CCSNe are usually only $0.04-0.2M_{\odot}$ for normal
CCSNe and $0.2-0.5M_{\odot }$ \citep[see e.g., Figure 8 of][]{Mazzali13} for
some Hypernovae whose kinetic energies were inferred to be $\gtrsim 10^{52}$
erg.\footnote{%
There is evidence that at least some hypernovae were powered by magnetars %
\citep{Wang2016b} and the required $^{56}$Ni mass and initial explosion
energy are significantly smaller than previously believed %
\citep{WangHan16,Wang2017}.} When we study an SLSNe whose peak luminosities $%
\gtrsim 10^{44}$ erg s$^{-1}$, the contribution of $0.1-0.5$ $M_{\odot}$ of $%
^{56}$Ni is significantly lower than the contribution from magnetars or the
ejecta-CSM interaction and can therefore be neglected %
\citep[e.g.,][]{Inse2013,Chen2015} since $0.1-0.5$ $M_{\odot}$ of $^{56}$Ni
will power a peak luminosity of $\sim 2\times 10^{42}-1\times 10^{43}$ erg~ s%
$^{-1}$ if the rise time is $\sim 18$ days and a lower peak luminosity for
longer rise time.\footnote{%
For the luminous SNe whose peak luminosity $\sim 5\times 10^{43}$ erg s$%
^{-1} $, the contribution from $^{56}$Ni cannot be neglected %
\citep{Wang2015b}.}

Hence, we neglect the contribution from $^{56}$Ni in our modeling, i.e., the
mass of $^{56}$Ni is set to be zero. In summary, the free parameters in this
model are $M_{\text{ej}}$, $B_{p}$, $P_{0}$, $v_{\text{sc0}}$, $%
\kappa_{\gamma}$, and $T_{\text{start}}$. To get the best-fitting parameters
and estimate the uncertainties of the parameters, we use the code developed
by \citet{Wang2016b} who incorporated MCMC approach into our revised
magnetar-powered model.

\section{Results}

\label{sec:res}

Using the magnetar-powered model and the MCMC approach, we find that the
light curves, temperature evolution, and velocity evolution reproduced by
the model are in excellent agreement with the observational data of LSQ14mo,
PS1-10awh, PS1-10bzj, SN 2010gx, SN 2011kf, and PS1-11ap, see Figure \ref%
{fig:fit1}. The light curves and temperature data of SN 2012il, PTF12dam,
PTF11rks, SN 2013dg, SSS120810, PTF10hgi, Gaia16apd, and DES13S2cmm can also
be explained by the model, but the velocities reproduced by the model are
larger or smaller than the data of these SLSNe, see Figure \ref{fig:fit2}.
DES13S2cmm is also in this group since it do not have velocity data. The
light curves of SN 2011ke, SN 2015bn, LSQ12dlf, DES14X3taz, and PS1-14bj can
also be well reproduced by the model, but both the temperature evolution and
velocity evolution of these SLSNe do not fit the theoretical curves well,
see Figure \ref{fig:fit3}. In these figures, the zero epochs all
refer to the peak times of the SLSN light curves.

The best-fitting parameters as well as the values of $\chi^2$/d.o.f are
listed in Table \ref{tbl:para} and their error bars can be appreciated by
inspecting Figure \ref{fig:corner}. Some parameters, e.g., $\kappa_{\gamma}$
and $v_{\text{sc0}}$\ of some SNe (see Figure \ref{fig:corner}), cannot be
tightly constrained and therefore only their $1\sigma $\ upper limits or
lower limits are presented in this table. From Table \ref{tbl:para}, we find
that the magnetars' initial periods and the magnetic field strengths are $%
1.2-8.3$ ms and $0.2-8.8\times 10^{14}$ G, respectively, consistent with
theoretical expectation and the results of previous modelings. The gamma-ray
opacity $\kappa_{\gamma }$ directly determining the magnitude of late-time
leakage effect is between 0.01 and 0.8 cm$^2$ g$^{-1}$. Assuming that $%
\kappa $ = 0.1 cm$^2$ g$^{-1}$, the masses of these SLSNe are between $%
1-27.6M_{\odot }$. \citet{Nich2015b} have fitted the light curves of 24
hydrogen-deficient SLSNe and found that the range of their ejecta masses is $%
3-30M_{\odot}$ if $\kappa =0.1$\thinspace cm$^2$ g$^{-1}$. Both the lower
limit and the upper limit of the mass range in our sample are smaller than
the values inferred by \citet{Nich2015b}. If we adopt a larger $\kappa$,
e.g., 0.2 cm$^2$ g$^{-1}$, the ejecta mass must be halved. Although the
acceleration effect must be obvious for less massive ejecta, the light
curves have only a slight change, i.e., the degeneracy between the $\kappa$
and $M_{\text{ej}}$ would not be broken.

The initial scale velocity $v_{\text{sc0}}$ of the ejecta of these SLSNe is
between $\sim $1,100 km s$^{-1}$ and $1.7\times 10^{4}$ km s$^{-1}$. Based
on the values of $M_{\text{ej}}$ and $v_{\text{sc0}}$, the initial KE $E_{%
\text{K0}}$ of these SLSNe can be calculated. In the beginning of the
expansion when the acceleration does not take place, the initial KE can be
calculated according to $E_{\text{K0}}=0.3M_{\text{ej}}v_{\text{sc0}}^{2}$
(see \citealt{Arn1982} and footnote 1 of \citealt{Wheeler15}) and listed in
Table \ref{tbl:derived parameters}. The initial rotational energy of the
corresponding magnetars and the accumulative fraction ($\eta $) of the
magnetars' rotational energy converted to the ejecta KE are also listed in
Table \ref{tbl:derived parameters}.

Figure \ref{fig:dis} shows the distributions of these best-fitting
parameters (initial periods $P_{0}$, magnetic strength $B_{p}$, ejecta
masses $M_{\text{ej}}$, initial scale velocities $v_{\text{sc0}}$, and the
gamma-ray opacity $\kappa _{\gamma }$), the derived parameter (initial KE $%
E_{\text{K0}}$), and the conversion fraction ($\eta$).

\section{Discussion}

\label{sec:dis}

In our adopted model, the initial velocity of the SN ejecta is also a free
parameter, rather than a measurable quantity. The magnetar wind would
accelerate the ejecta, sweep up the material into a (thin) shell and produce
a bubble between the center and the shell. We fit the observed photospheric
velocities $v_{\text{ph}}$ (even though they are always smaller than the
scale velocities $v_{\text{sc}}$) and find the theoretical velocities of the
ejecta of 6 SLSNe are in excellent agreement with the observations. However,
we would point out that the velocity data of the remaining SLSNe in our
sample cannot be well reproduced by the model.

In most of the previous studies, the KE $E_{\text{K}}$ was a constant since
the scale velocity $v_{\text{sc}}$ is fixed to be a constant ($v_{\text{sc}%
}=v_{\text{sc0}}$), then the inferred KE is equivalent to the initial KE ($%
E_{\text{K}}=E_{\text{K0}}$). This assumption would cause two problems.

First, if we neglect the acceleration effect from the magnetar wind, the
ejecta expansion is homologous ($v_{\text{sc}}(x)\propto x$, where $x$ is
the distance between the arbitrary point in the ejecta to the center), and
the KE of the ejecta is $E_{\text{K}}=0.3M_{\text{ej}}v_{\text{sc}}^2$.
According to this formula, we find that the initial KEs of almost all SLSNe
in the literature are $\gtrsim 5\times 10^{51}$ erg if the acceleration
effect is neglected. Most SLSNe might be CCSNe and their initial KE should
be given by neutrinos coming from proto-neutron stars. The multi-dimensional
simulations of neutrino-driven SNe find that the upper limit of the KE
provided by neutrinos is $\sim 2.0\times 10^{51}$ erg or $\sim 2.5\times
10^{51}$ erg \citep{Ugl2012,Ertl2016,Mul2016,Suk2016}. Even if we adopt the
looser upper limit ($\sim 2.5\times 10^{51}$ erg), it is still significantly
smaller than the inferred KE ($\gtrsim 5\times 10^{51}$ erg) of the SLSNe.
It seems evident that some other mechanisms are required to provide the
additional initial KE.

To solve this problem, one might assume that the energy injection process
must be divided into two steps: (1) the magnetar releases $\gtrsim 3\times
10^{51}$ erg of the rotational energy at very short internal (``explosive
injection"), and (2) the magnetar continuously injects its rotational energy
to SN ejecta (``continuous injection"). As pointed out by \citet{Ioka2016},
this two-step injection scenario needs an exotic behavior: the magnetic
field should be initially large, resulting in a fast energy injection for a
short duration, and then it may decay rapidly between the explosive
injection and continuous injection.

Second, the initial rotational energy of a magnetar with an initial period $%
P_{0}$ is $\frac{1}{2}I\Omega_{0}^2\simeq 2\times 10^{52}\left(I/10^{45} {~%
\text{g}~\text{cm}}^2\right) \left({P_{0}}/{1~\text{ms}}\right) ^{-2}$ erg.
Here $I$\ is the moment of inertia of the magnetar and $\Omega_{0}$\ is its
initial angular velocity. If $P_{0}=1-5$ ms, the initial rotational energy
is $\simeq 1-20\times 10^{51}$ erg. Calculations %
\citep{Kas2010,Woos2010,Wang2016b} have indicated that a fraction of the
initial rotational energy of a magnetar would be converted to the KE of SNe.
\footnote{%
For example, \citet{Woos2010} demonstrated that a magnetar with $P_{0}$ =
4.5 ms and $B_{p}=1\times 10^{14}$ G would convert 40\% of its initial
rotational energy to the KE of the SN ejecta.} This huge amount of energy
and its acceleration effect cannot be neglected.

Our modelings based on the magnetar-powered model \citep{Wang2016b} can
simultaneously solve these two problems by taking into account the
acceleration effect of the magnetar wind. We find that $\sim 19-97$\% of the
initial rotational energy of the magnetars have been converted to the KE of
the ejecta (see Table \ref{tbl:derived parameters} and Figure \ref{fig:fra}
for some of these SLSNe) and the initial KE of 15 SLSNe in our sample are
smaller than $2.5\times 10^{51}$ erg (see Table \ref{tbl:derived parameters}%
) provided by the neutrino-driven mechanism for CCSNe. The additional KE is
provided by the magnetar wind which accelerates the ejecta.

Besides, LSQ14mo, SN 2015bn and DES13S2cmm have initial KE (slightly) larger
than $\sim 2.5\times 10^{51}$ erg but the initial KE of these SLSNe can be
halved to be $\sim 1.37\times 10^{51}$, $\sim 1.41\times 10^{51}$, and $\sim
2.28\times 10^{51}$ erg if the value of $\kappa$ doubled ($\kappa =0.2$ ~cm$%
^{-2}$ g$^{-1}$) so that their ejecta masses can be halved. Therefore, these
three SLSNe can also be explained by the magnetar model. The only SLSN whose
KE cannot be explained by the magnetar-powered model is DES14X3taz since its
initial KE is $4.51\times 10^{52}$ erg (if $\kappa =0.1$ cm$^{-2}$ ~g$^{-1}$%
) or $2.25\times 10^{52}$ erg (if $\kappa =0.2$ cm$^{-2}$ g$^{-1}$),
significantly larger than that can be provided by the neutrinos.

All in all, in our model, the neutrinos provide the initial KE of the
ejecta, and the magnetar accelerates the ejecta so that they have large
final KE ($\gtrsim 5\times 10^{51}$ erg). Thus the difficulty of explaining
the origin of the KE of the SLSNe can be solved by taking into account the
acceleration effect. \footnote{%
The models proposed by \citet{Kas2010} and \citet{Woos2010} also took into
account the acceleration effect and can also solve the problem associated
with the initial KE. However, these models neglected the leakage effect.}

Figure \ref{fig:etavsb} shows the conversion fraction $\eta$ versus magnetic
strength $B_{p}$. The correlation coefficient between these two quantities
is $R=0.715$, which means larger $B_{p}$ results in a larger $\eta$. \cite%
{Wang2016a} demonstrated that the conversion fraction $\eta$ is high if the
spin-down timescale of the magnetar is short compared to the diffusion
timescale. It is therefore expected that the stronger $B_{p}$, the higher $%
\eta$ tends to be, although the other factors, e.g. $M_{\text{ej}}$, $P_{0}$%
, will cause some scatters in this relation. If $B_{p}$ is increased further
to $\sim 10^{16}$ G, we can expect that most of the magnetar's rotational
energy will be converted to ejecta's KE and the SNe will become dimmer. This
is precisely what is seen for broad-lined type Ic SNe (SNe Ic-BL) by \cite%
{WangHan16}, who showed evidence that SNe Ic-BL 1998bw and 2002ap were
powered by magnetars \citep{Wang2016b}. This implies a continuous spectrum
of magnetar-powered SNe.

The inferred gamma-ray opacity $\kappa_{\gamma}$ is $\sim 0.01-0.82$ cm$%
^{-2} $ ~g$^{-1}$. \citet{Kot2013} has demonstrated that the values of $%
\kappa_{\gamma}$ must be between approximately 0.01 and 0.2 cm$^2$ g$^{-1}$
if the emission is dominated by gamma photons, and between approximately 0.2
and $10^{4}$ ~cm$^2$ g$^{-1}$ if the emission is dominated by X-ray photons.
In our sample, the values of $\kappa_{\gamma}$ of PTF12dam and SSS 120810
reach the lower limit, $\sim $ 0.01 cm$^2$ g$^{-1}$, proposed by the
theoretical prediction (see also \citealt{Chen2015} for PTF12dam),
suggesting that the energy of the gamma-ray radiation from the magnetars
associated with these SLSNe must be higher than $3\times 10^{7}$ eV = 30
MeV. Such high-energy photons from magnetars have already been observed %
\citep{Hester08, Buhler14} and can be explained by theoretical modelings %
\citep{Metzger14,Murase15,WangDai16}.

For SLSNe in Figure \ref{fig:fit2}, bolometric luminosity ($L$) and
temperature ($T$) evolution can be well fitted but the evolution of
photospheric velocity ($v_\text{ph}$) cannot be fitted. This is strange
since $L=4\pi \sigma T^4R_\text{ph}^2 =4\pi \sigma T^4{(\int_{v_\text{ph0}%
}^{v_\text{ph}}v_\text{ph}dt)}^2$ ($R_\text{ph}=\int_{v_\text{ph0}}^{v_\text{%
ph}}v_\text{ph}dt$ is the photospheric radius). The reason for this might be
that the error bars of the observational data are too small.
Observationally, the velocities inferred from different elements are rather
different (e.g., see Figure 7 of \citealt{Tad2016}), so the error bars of
the velocity data might be larger than that presented in the Figure \ref%
{fig:fit2}. To clarify this issue, more dedicated studies are required.

Some of the SLSNe in Figure \ref{fig:fit3} show flattening (SN 2011ke, SN
2015bn, PS1-11ap, PS1-14bj) in the late-time temperature evolution. This
flattening may be caused by recombination, which is not considered in our
adopted model. The most extreme case is PS1-14bj, whose temperature
evolution is rather flat at very beginning, and even slightly increasing
with time. Another peculiar case is DES14X3taz, whose temperature evolution
is untypical in the SN sample. We note that the SLSNe mentioned above
(PS1-14bj, SN 2015bn and DES14X3taz) are among the four SLSNe that are not
well fitted by our model. These peculiar SLSNe deserve more investigations.
The large reduced $\chi^2$ of SN 2013dg (among the four largest reduced $%
\chi^2$), on the other hand, is caused by the small errors in temperature
measurement because its light-curve and velocity fitting quality is similar
to PTF11rks.

\section{Conclusions}

\label{sec:con}

Detailed studies of SLSNe in the last decade have revealed many important
observational properties and given some crucial clues to understanding the
energy-source models and the explosion mechanisms of SLSNe. In the last
several years, most SLSNe discovered are type I whose light curves and
spectra are diverse and complex \citep[e.g.,][]{Inse2017}.

Using the magnetar-powered model and the MCMC approach, we fit the light
curves, temperature evolution, and photospheric velocity evolution of 19
SLSNe I. We get rather good fits for 7 events ($\chi^2$/d.o.f = $0.24-0.96$)
and good fits for other 7 events ($\chi^2$/d.o.f = $1.37-3.13$), suggesting
that these SLSNe can be explained by the magnetar model. Four events cannot
be well fitted by this model ($\chi^2$/d.o.f = $4.32-6.83$), suggesting that
these four events must be further studied.

The parameters determined by the MCMC code are as follows. The values of the
initial period of the magnetars supposed to power these SLSNe are $1.2-8.3$
ms; the values of the magnetic strength of the magnetars are $0.2-8.8\times
10^{14}$ G; the masses of the ejecta of these SLSNe are $1-27.6 M_{\odot}$;
the gamma-ray opacity are $0.01-0.82$ cm$^2$ g$^{-1}$.

More importantly, we take into account the acceleration effect and let the
initial velocity of the ejecta be a free parameter and find that the initial
KE of most SLSNe in our sample are (significantly) smaller than the upper
limit ($\sim 2.5\times 10^{51}$ erg) of the KE provided by the
neutrino-driven mechanism for CCSNe, indicating that our modelings are
self-consistent and do not need any exotic assumption (e.g., two-step
injections from the magnetars) to explain the origin of the ejecta kinetic
energy of these SLSNe.

Our modeling shows that $\sim $ 19$-$97\% of the initial rotational energy
of the magnetars is converted to the KE of the SNe ejecta. This acceleration
effect is especially important in the SLSNe that require magnetars with
initial periods $P_{0}\sim 1-5$ ms, since the initial rotational energy of
these magnetars is $\sim 1-20\times 10^{51}$~erg and would convert $\sim
0.2-20\times 10^{51}$~erg to KE of the SLSNe ejecta.

By combining these two results, we demonstrate that the KE acquired from the
rotational energy dissipated via magnetic dipole radiation can naturally
provide a considerable amount of the KE for the SN ejecta, and the
difficulty of explaining the KE of the SLSNe whose KE are usually
(significantly) larger than $2\times 10^{51}$ erg can be solved.

Understanding the nature of SLSNe is one of the most challenging questions
in astrophysics. Their explosion mechanisms and energy sources are still
ambiguous. Our results provide some new and important clues related to these
problems. To clarify these important issues, more observations and
theoretical work are needed to be done.

\acknowledgments We thank the referee for very constructive suggestions
which have allowed us to improve this manuscript significantly. This work
was supported by the National Basic Research Program (\textquotedblleft
973\textquotedblright\ Program) of China (grant no. 2014CB845800) and the
National Natural Science Foundation of China (grants no. 11573014, U1331202,
11533033, 11422325, 11373022, and 11673006).

\clearpage

\begin{table*}[tbph]
\caption{SLSNe in our sample}
\label{tbl:sample}
\begin{center}
\begin{tabular}{llll}
\hline\hline
SN name & Redshift ($z$) & Peak Luminosity & Reference \\
&  & ($10^{44}$ erg s$^{-1}$) &  \\ \hline
DES13S2cmm & 0.663 & 0.63 & \citep{Papa2015} \\
DES14X3taz & 0.608 & 2.20 & \citep{Smit2016} \\
Gaia16apd & 0.102 & 3.01 & \citep{Kan2016} \\
LSQ12dlf & 0.25 & 1.03 & \citep{Nich2014} \\
LSQ14mo & 0.256 & 1.09 & \citep{Chen2016} \\
PS1-10awh & 0.908 & 2.41 & \citep{Chom2011} \\
PS1-10bzj & 0.65 & 1.07 & \citep{Lunn2013} \\
PS1-11ap & 0.524 & 0.72 & \citep{McC2014} \\
PS1-14bj & 0.521 & 0.46 & \citep{Lunn2016} \\
PTF10hgi & 0.099 & 0.35 & \citep{Inse2013} \\
PTF11rks & 0.193 & 0.47 & \citep{Inse2013} \\
PTF12dam & 0.108 & 1.17 & \citep{Nich2013} \\
SN 2010gx & 0.231 & 0.91 & \citep{Inse2013} \\
SN 2011ke & 0.143 & 0.81 & \citep{Inse2013} \\
SN 2011kf & 0.245 & 3.72 & \citep{Inse2013} \\
SN 2012il & 0.175 & 0.80 & \citep{Inse2013} \\
SN 2013dg & 0.26 & 0.74 & \citep{Nich2014} \\
SN 2015bn & 0.114 & 2.42 & \citep{Nich2016a} \\
SSS 120810 & 0.17 & 1.04 & \citep{Nich2014} \\ \hline
\end{tabular}%
\end{center}
\end{table*}

\clearpage

\begin{table*}[tbph]
\caption{Best-fit parameters for our sample of SLSNe.}
\label{tbl:para}
\begin{center}
\begin{tabular}{llllllll}
\hline\hline
SN name & $P_{0}$ & $B_{p}$ & $M_{\text{ej}}$ & $\kappa _{\gamma }$ & $v_{%
\text{sc0}}$ & $T_{\text{start}}$ & $\chi^2/$dof \\
& (ms) & ($10^{14}$ G) & ($M_{\odot}$) & (cm$^2$ g$^{-1}$) & (km s$^{-1}$) &
(days) &  \\ \hline
DES13S2cmm & 5.5 $_{-0.07}^{+0.07 }$ & 1.8 $_{-0.13}^{+0.14 }$ & 7.2 $%
_{-0.94 }^{+1.14 }$ & 0.33 $_{-0.18}^{+0.18 }$ & 10286 $_{ -453 }^{+ 493 }$
& -33.0 $_{ -1.1 }^{+ 1.1 }$ & 3.12 \\
DES14X3taz & 1.3 $_{ -0.28 }^{+ 0.20 }$ & 0.2 $_{ -0.05 }^{+ 0.08 }$ & 24.5 $%
_{ -3.60 }^{+ 4.83 }$ & $\leq$0.06 & 17517 $_{ -1213 }^{+ 1267 }$ & -40.1 $%
_{ -1.0 }^{+ 1.2 }$ & 4.32 \\
Gaia16apd & 2.1 $_{-0.13 }^{+0.18 }$ & 2.2 $_{ -0.10 }^{+ 0.13 }$ & 3.2 $_{
-0.20 }^{+ 0.21 }$ & 0.57 $_{-0.32 }^{+ 0.21 }$ & 8618 $_{ -1204 }^{+ 1002 }$
& -35.9 $_{ -0.6 }^{+ 0.9 }$ & 0.64 \\
LSQ12dlf & 1.3 $_{ -0.09 }^{+ 0.33 }$ & 5.5 $_{ -0.23 }^{+ 0.14 }$ & 14.5 $%
_{ -1.31 }^{+ 1.20 }$ & 0.23 $_{ -0.18 }^{+ 0.20 }$ & $\leq$2226 & -21.2 $_{
-0.6 }^{+ 0.4 }$ & 3.13 \\
LSQ14mo & 3.0 $_{ -0.82 }^{+ 0.76 }$ & 6.9 $_{ -3.89 }^{+ 0.68 }$ & 6.8 $_{
-0.62 }^{+ 2.59 }$ & 0.26 $_{ -0.17 }^{+ 0.19 }$ & 8192 $_{ -1929 }^{+ 1959
} $ & -21.1 $_{ -6.6 }^{+ 0.9 }$ & 0.40 \\
PS1-10awh & 2.5 $_{ -1.01 }^{+ 1.15 }$ & 0.5 $_{ -0.36 }^{+ 1.17 }$ & 1.0 $%
_{ -0.39 }^{+ 1.79 }$ & 0.10 $_{ -0.07 }^{+ 0.07 }$ & 13383 $_{ -425 }^{+
427 }$ & -24.8 $_{ -0.5 }^{+ 0.5 }$ & 0.44 \\
PS1-10bzj & 1.2 $_{ -0.24 }^{+ 0.41 }$ & 7.8 $_{ -0.49 }^{+ 0.40 }$ & 12.2 $%
_{ -2.45 }^{+ 3.35 }$ & 0.30 $_{ -0.20 }^{+ 0.20 }$ & 1740 $_{ -1478 }^{+
1735 }$ & -18.6 $_{ -0.5 }^{+ 0.3 }$ & 0.24 \\
PS1-11ap & 3.9 $_{ -0.08 }^{+ 0.09 }$ & 2.1 $_{ -0.06 }^{+ 0.06 }$ & 6.1$_{
-0.40 }^{+ 0.46 }$ & 0.22 $_{ -0.15 }^{+ 0.25 }$ & $\leq$3536 & -48.6 $_{
-0.9 }^{+ 0.9 }$ & 0.53 \\
PS1-14bj & 2.7 $_{ -0.14 }^{+ 0.16 }$ & 0.4 $_{ -0.06 }^{+ 0.08 }$ & 27.6 $%
_{ -2.71 }^{+ 2.81 }$ & $\leq$0.08 & 1158$_{ -594 }^{+ 450 }$ & -172.6 $_{
-6.4 }^{+ 6.4 }$ & 6.83 \\
PTF10hgi & 8.3 $_{ -0.25 }^{+ 0.22 }$ & 3.0 $_{ -0.17 }^{+ 0.16 }$ & 1.2 $_{
-0.16 }^{+ 0.38 }$ & 0.19 $_{ -0.09 }^{+ 0.21 }$ & $\leq$4112 & -5.3 $_{
-2.2 }^{+ 1.9 }$ & 1.37 \\
PTF11rks & 6.9 $_{ -0.89 }^{+ 0.68 }$ & 8.8 $_{ -0.87 }^{+ 1.04 }$ & 3.5 $_{
-0.49 }^{+ 0.57 }$ & $\geq$0.15 & 8044 $_{ -574 }^{+ 576 }$ & 1.3 $_{ -0.5
}^{+ 0.5 }$ & 2.94 \\
PTF12dam & 2.7 $_{ -0.13 }^{+ 0.10 }$ & 0.8 $_{ -0.20 }^{+ 0.17 }$ & 12.0 $%
_{ -2.55 }^{+ 3.48 }$ & 0.01 $_{ -0.005 }^{+ 0.006 }$ & $\leq$2345 & -69.9 $%
_{ -3.6 }^{+ 3.1 }$ & 0.39 \\
SN 2010gx & 1.3 $_{ -0.20 }^{+ 1.07 }$ & 8.5 $_{ -1.18 }^{+ 0.38 }$ & 12.8 $%
_{ -1.29 }^{+ 1.28 }$ & 0.04 $_{ -0.02 }^{+ 0.24 }$ & 4233 $_{ -3902 }^{+
4438 }$ & -1.8 $_{ -1.0 }^{+ 0.8 }$ & 0.96 \\
SN 2011ke & 2.6 $_{ -0.17 }^{+ 0.22 }$ & 7.1 $_{ -0.15 }^{+ 0.15 }$ & 6.4 $%
_{ -0.47 }^{+ 0.50 }$ & 0.71 $_{ -0.24 }^{+ 0.20 }$ & 1816 $_{ -1045 }^{+
2270 }$ & 9.7 $_{ -0.6 }^{+ 0.5 }$ & 3.46 \\
SN 2011kf & 2.1 $_{ -0.15 }^{+ 0.16 }$ & 5.2 $_{ -0.12 }^{+ 0.11 }$ & 2.1 $%
_{ -0.15 }^{+ 0.16 }$ & 0.82 $_{ -0.14 }^{+ 0.12 }$ & $\leq$3704 & 0.5 $_{
-0.5 }^{+ 0.4 }$ & 1.91 \\
SN 2012il & 5.9 $_{ -0.15 }^{+ 0.15 }$ & 4.6 $_{ -0.05 }^{+ 0.05 }$ & 1.6 $%
_{ -0.15 }^{+ 0.17 }$ & $\geq$0.4 & $\leq$3595 & -2.3 $_{ -0.7 }^{+ 0.6 }$ &
2.47 \\
SN 2013dg & 4.4 $_{ -0.67 }^{+ 0.59 }$ & 1.2 $_{ -0.36 }^{+ 0.53 }$ & 1.2 $%
_{ -0.27 }^{+ 0.31 }$ & $\geq$0.15 & $\leq$4072 & -30.5 $_{ -1.1 }^{+ 1.4 }$
& 5.52 \\
SN 2015bn & 2.3 $_{ -0.10 }^{+ 0.05 }$ & 0.5 $_{ -0.05 }^{+ 0.03 }$ & 10.1 $%
_{ -0.20 }^{+ 0.37 }$ & 0.03 $_{ -0.007 }^{+ 0.008 }$ & 6798 $_{ -113 }^{+
150 }$ & -65.1 $_{ -1.5 }^{+ 1.1 }$ & 6.15 \\
SSS 120810 & 2.5 $_{ -0.22 }^{+ 0.27 }$ & 3.7 $_{ -0.22 }^{+ 0.19 }$ & 2.9 $%
_{ -1.42 }^{+ 1.45 }$ & 0.01 $_{ -0.006 }^{+ 0.007 }$ & 10961 $_{ -2435 }^{+
1728 }$ & -30.2 $_{ -0.4 }^{+ 0.2 }$ & 1.76 \\ \hline
\end{tabular}%
\end{center}
\par
\textbf{Note.} The uncertainties are $1\sigma $ errors.
\end{table*}

\clearpage

\begin{table*}[tbph]
\caption{Derived physical parameters}
\label{tbl:derived parameters}
\begin{center}
\begin{tabular}{lllll}
\hline\hline
SN name & $E_{K0}$\tablenotemark{a} & $E_{p0}$\tablenotemark{b} & $\eta $%
\tablenotemark{c} & ($P_{0}$, $B_{p}$) \\
& ($10^{51}$ erg) & ($10^{51}$ erg) &  & (ms, $10^{14}$ G) \\ \hline
DES13S2cmm & 4.55 & 0.65 & 0.89 & (5.5, 1.8) \\
DES14X3taz & 45.1 & 11.8 & 0.27 & (1.3, 0.2) \\
Gaia16apd & 1.99 & 3.77 & 0.46 & (2.1, 2.2) \\
LSQ12dlf & $0.1-$0.43 & 10.92 & 0.97 & (1.3, 5.5) \\
LSQ14mo & 2.74 & 2.14 & 0.86 & (3.0, 6.9) \\
PS1-10awh & 1.08 & 3.15 & 0.39 & (2.5, 0.5) \\
PS1-10bzj & 0.22 & 13.25 & 0.97 & (1.2, 7.8) \\
PS1-11ap & $0.1-$0.45 & 1.30 & 0.53 & (3.9, 2.1) \\
PS1-14bj & 0.22 & 2.81 & 0.23 & (2.7, 0.4) \\
PTF10hgi & $0.1-$0.12 & 0.28 & 0.27 & (8.3, 3.0) \\
PTF11rks & 1.37 & 0.42 & 0.68 & (6.9, 8.8) \\
PTF12dam & $0.1-$0.39 & 2.76 & 0.38 & (2.7, 0.8) \\
SN 2010gx & 1.38 & 11.82 & 0.97 & (1.3, 8.5) \\
SN 2011ke & 0.13 & 2.92 & 0.90 & (2.6, 7.1) \\
SN 2011kf & $0.1-$0.17 & 4.49 & 0.80 & (2.1, 5.2) \\
SN 2012il & $0.1-$0.12 & 0.57 & 0.52 & (5.9, 4.6) \\
SN 2013dg & $0.1-$0.12 & 1.03 & 0.19 & (4.4, 1.2) \\
SN 2015bn & 2.80 & 3.87 & 0.20 & (2.3, 0.5) \\
SSS120810 & 2.12 & 3.24 & 0.83 & (2.5, 3.7) \\ \hline
\end{tabular}%
\end{center}
\par
a. $E_{K0}$ is the initial kinetic energy of the SLSNe and its lower-limit
has been set to be $1 \times 10^{50}$ erg.
\par
b. $E_{p0} \simeq 2 \times 10^{52}\left({P_0}/{1~\text{ms}}\right)^{-2}$~erg
is the initial rotational energy of the magnetars proposed to power the
SLSNe.
\par
c. $\eta$ is the accumulative fraction of the rotational energy of the
magnetar converted to the kinetic energy of these SLSNe.
\end{table*}

\clearpage

\begin{figure}[tbph]
\begin{center}
\includegraphics[width=0.4\textwidth,angle=0]{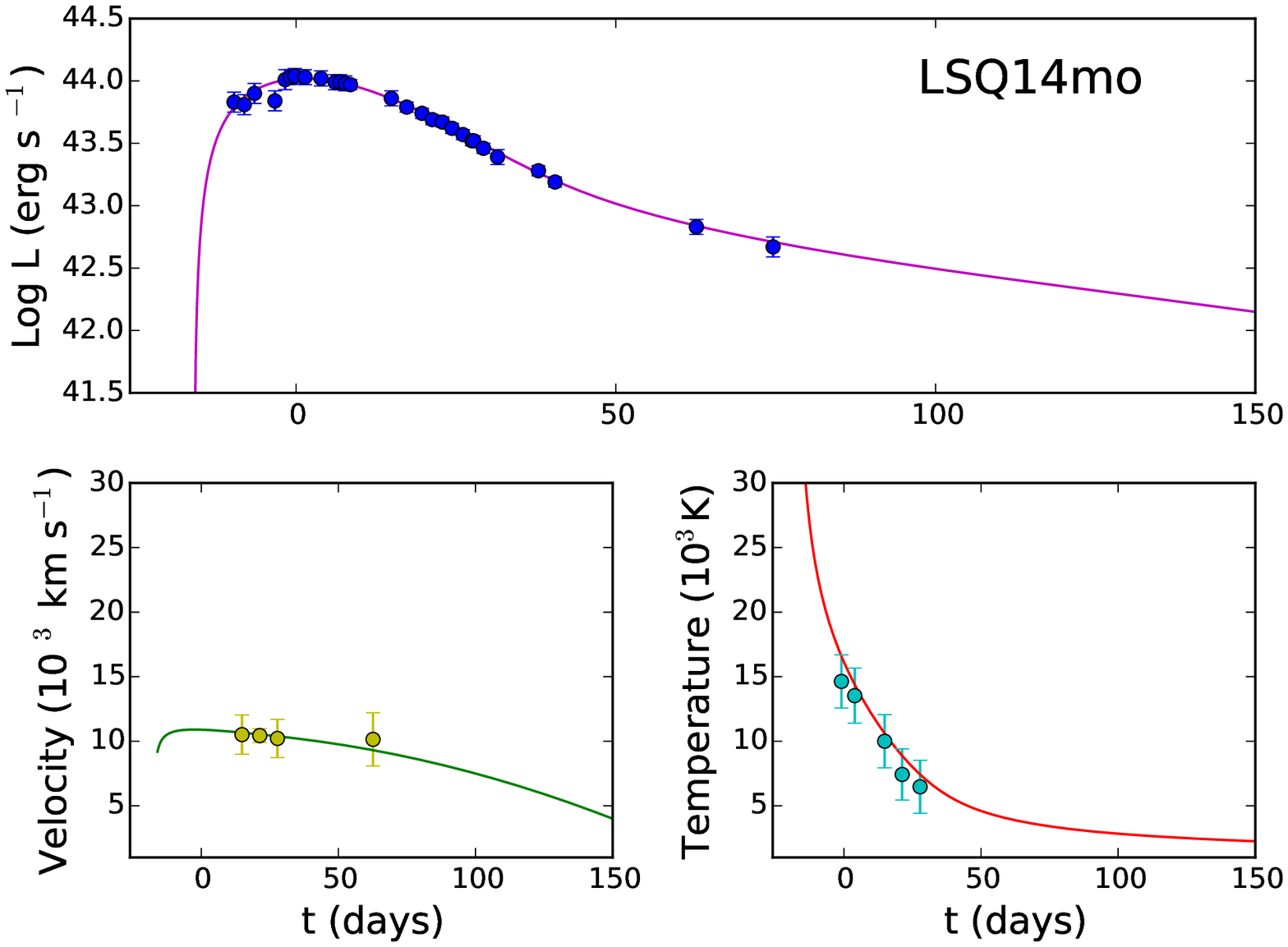}
\includegraphics[width=0.4\textwidth,angle=0]{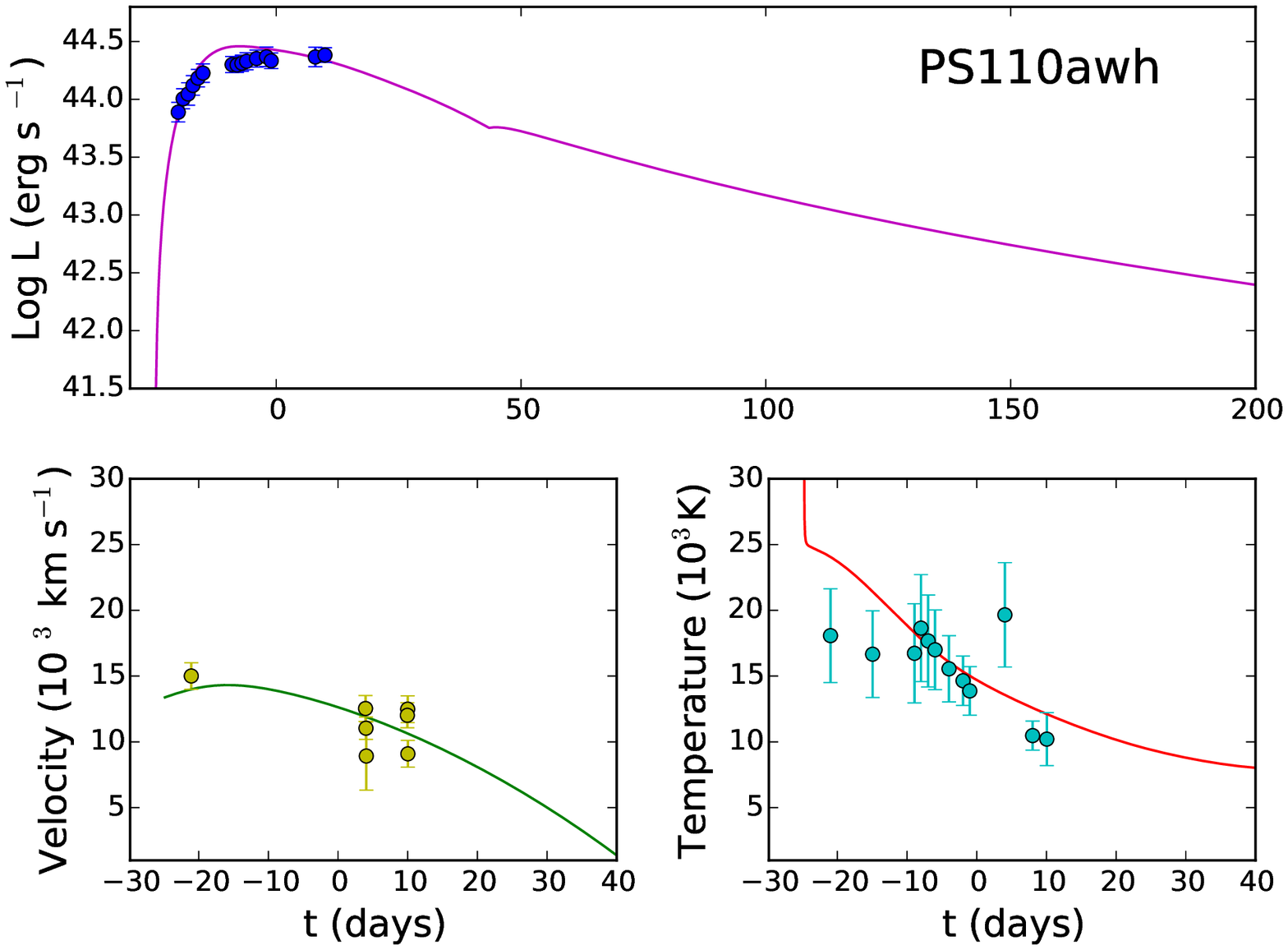}
\par
\includegraphics[width=0.4\textwidth,angle=0]{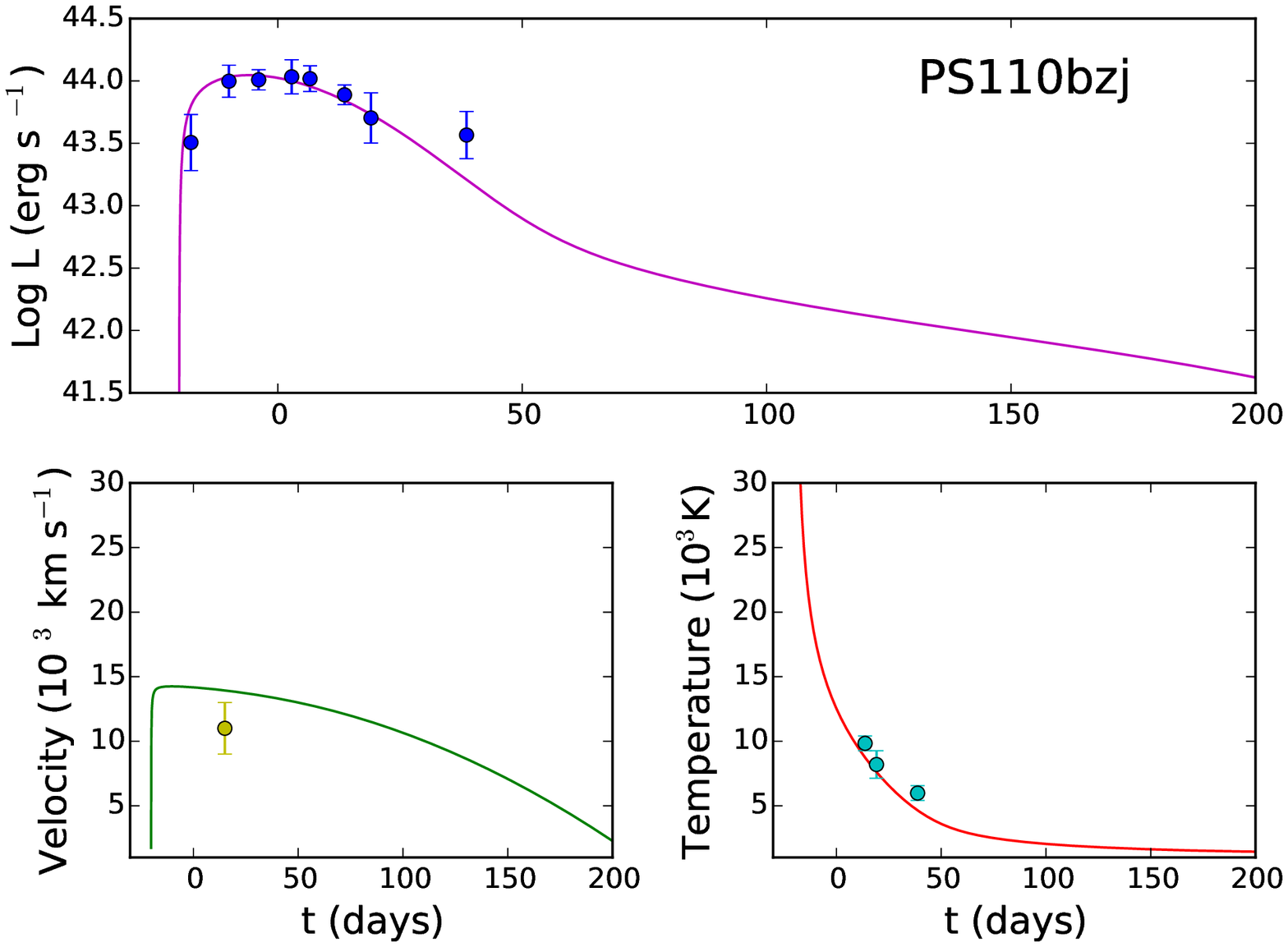}
\includegraphics[width=0.4\textwidth,angle=0]{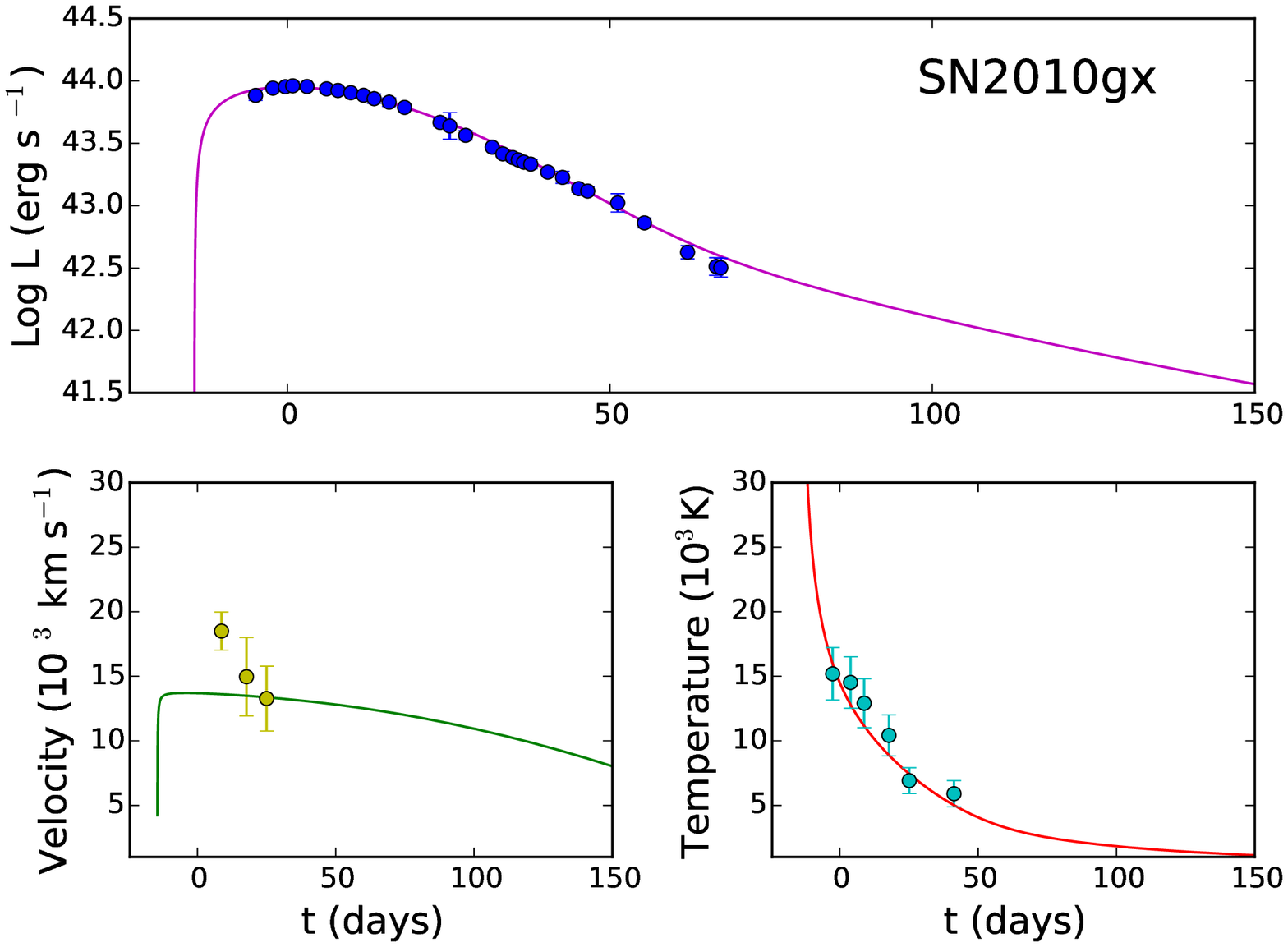}
\par
\includegraphics[width=0.4\textwidth,angle=0]{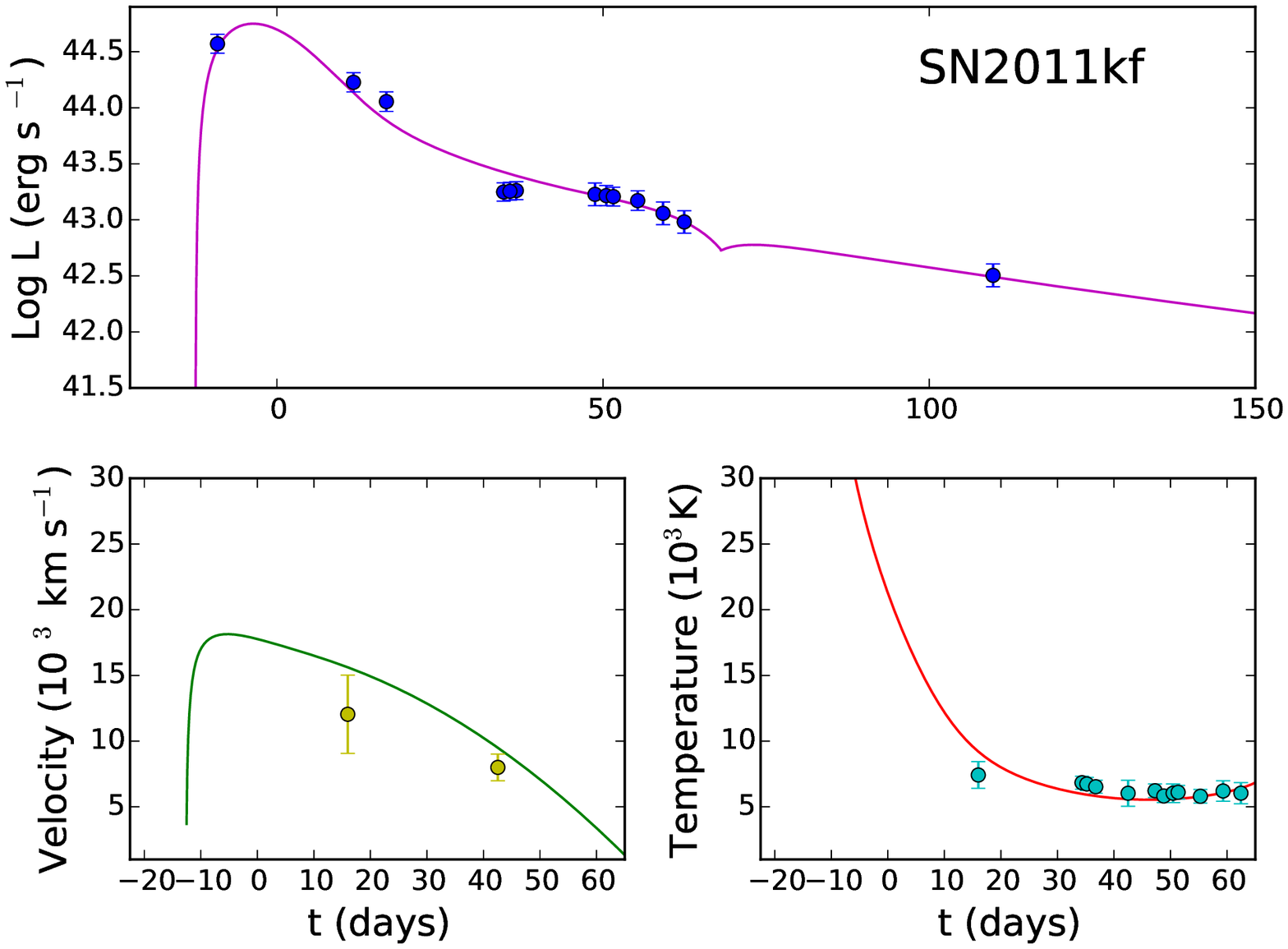}
\includegraphics[width=0.4\textwidth,angle=0]{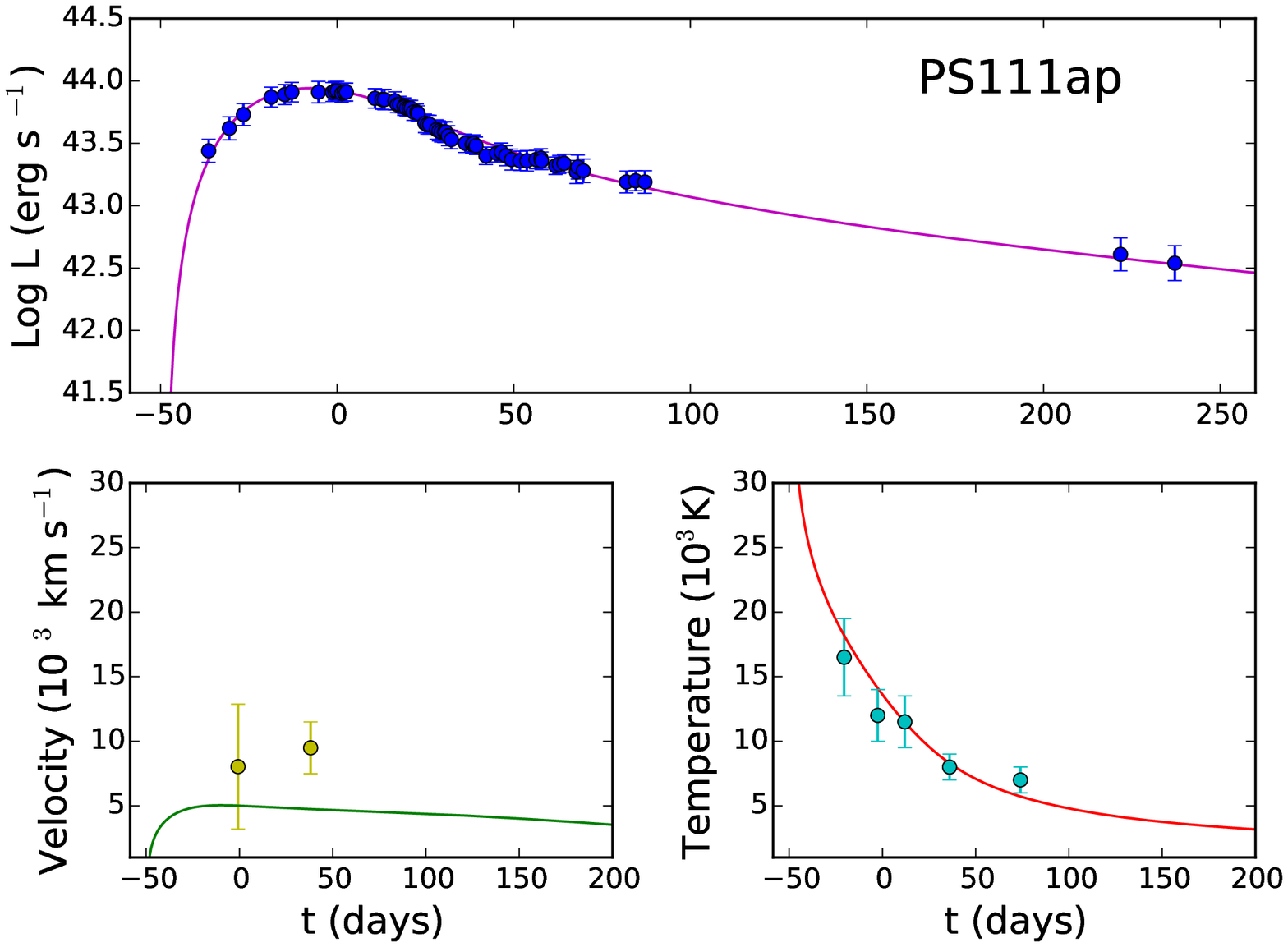}
\end{center}
\caption{Model fits for light curves, temperature evolution, and velocity
evolution of LSQ14mo, PS1-10awh, PS1-10bzj, SN 2010gx, SN 2011kf, and
PS1-11ap. Parameters are shown in Table \protect\ref{tbl:para}.}
\label{fig:fit1}
\end{figure}

\clearpage

\begin{figure}[tbph]
\begin{center}
\includegraphics[width=0.4\textwidth,angle=0]{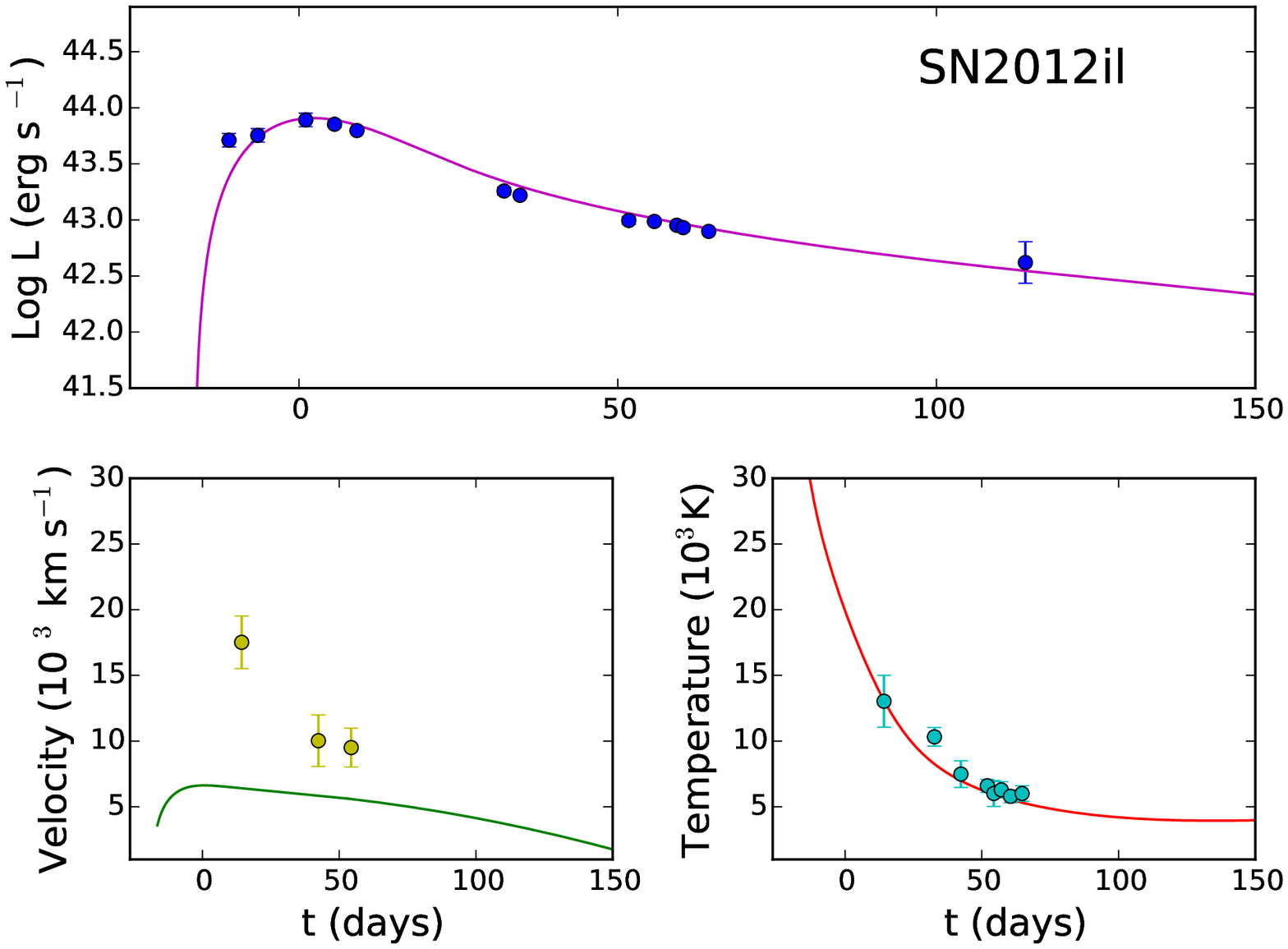}
\includegraphics[width=0.4\textwidth,angle=0]{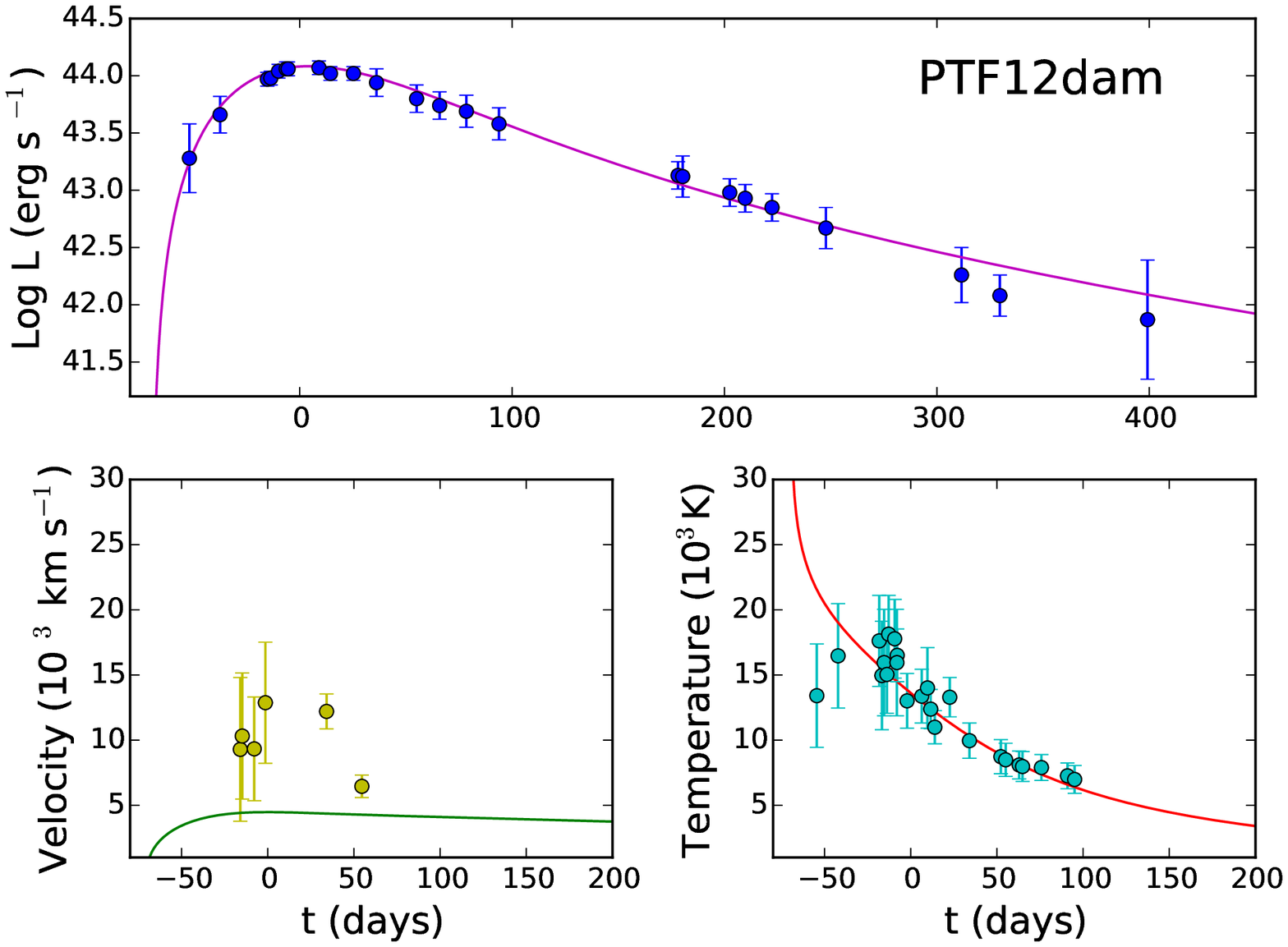}
\par
\includegraphics[width=0.4\textwidth,angle=0]{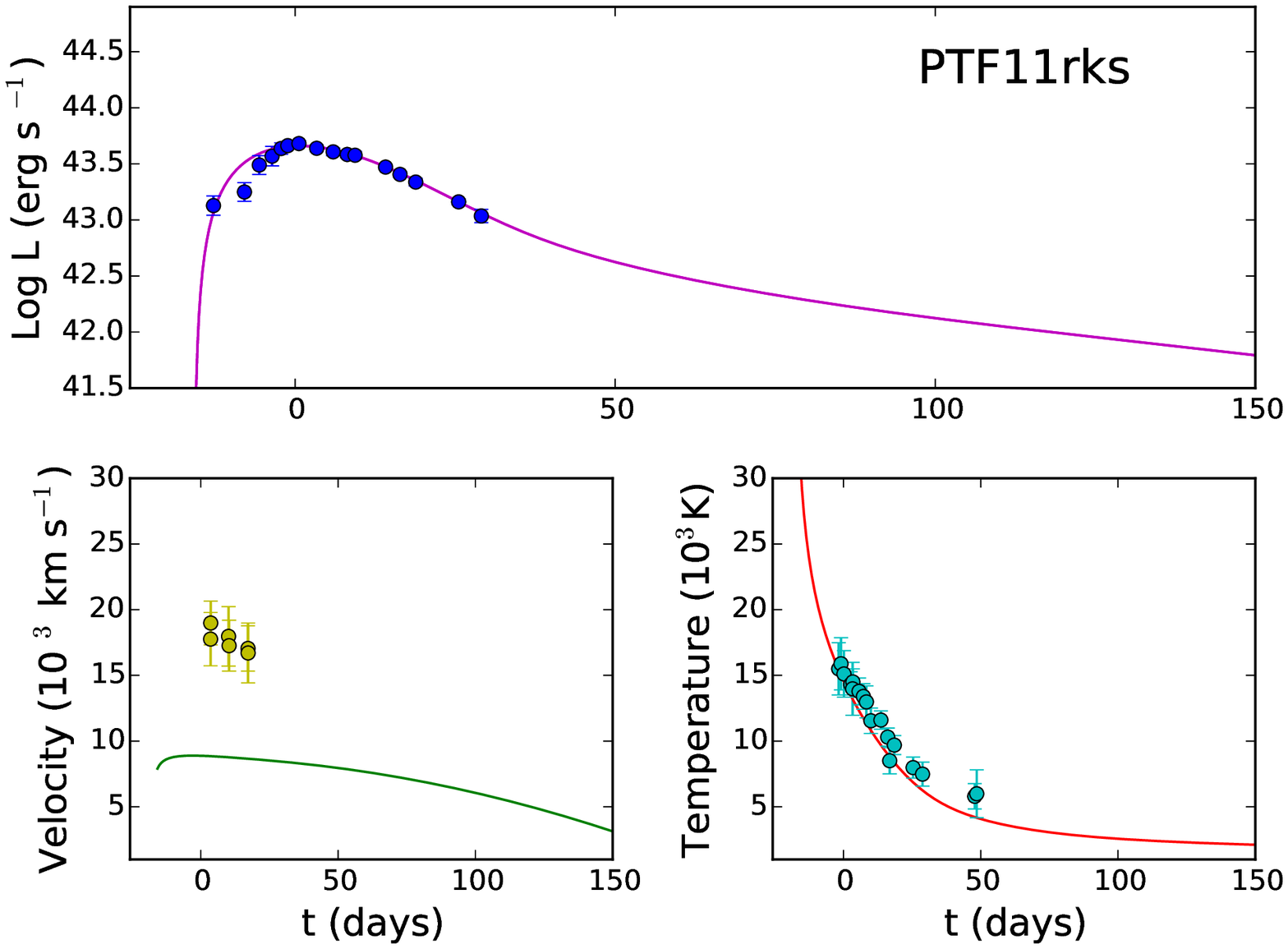}
\includegraphics[width=0.4\textwidth,angle=0]{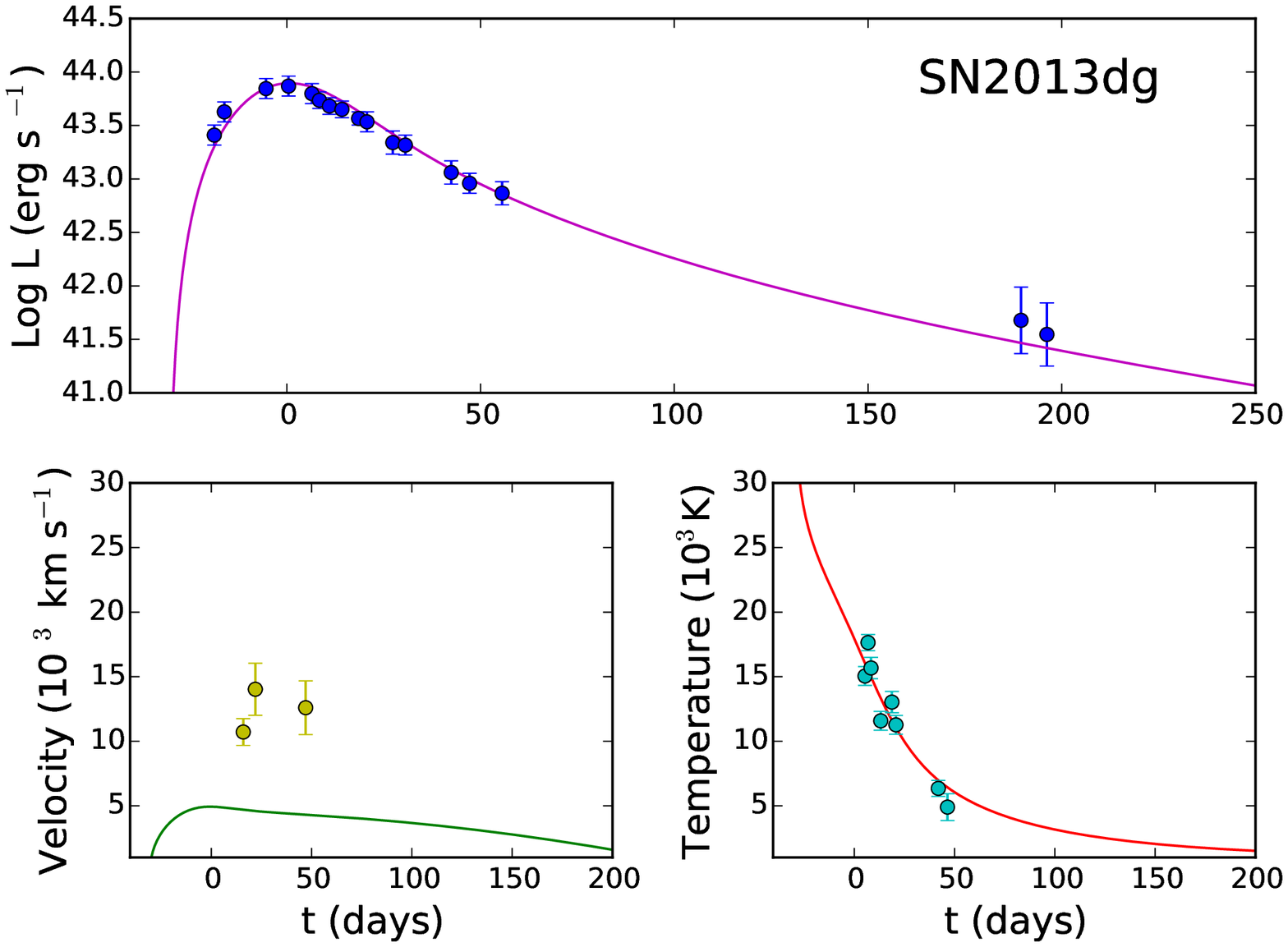}
\par
\includegraphics[width=0.4\textwidth,angle=0]{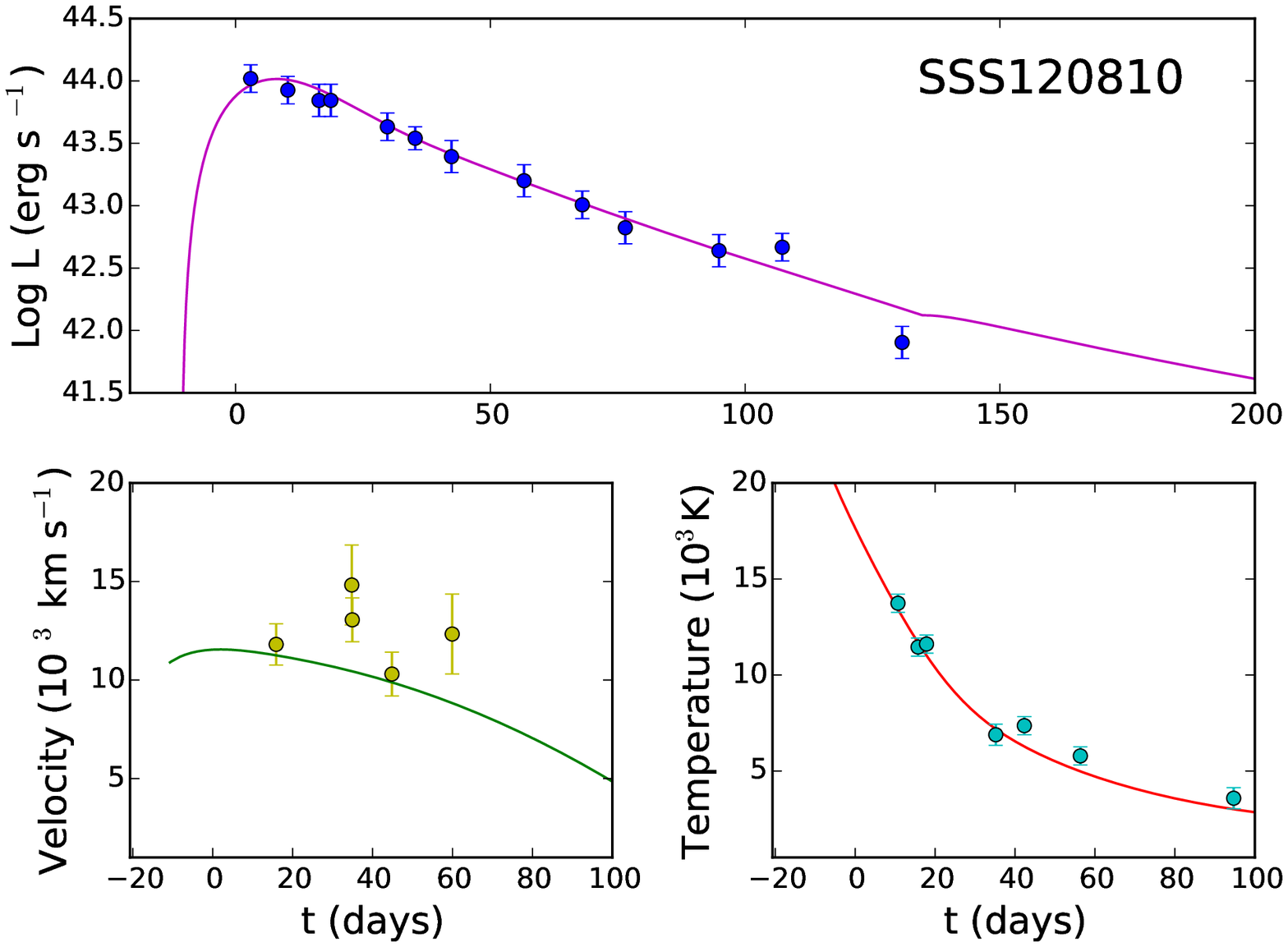}
\includegraphics[width=0.4\textwidth,angle=0]{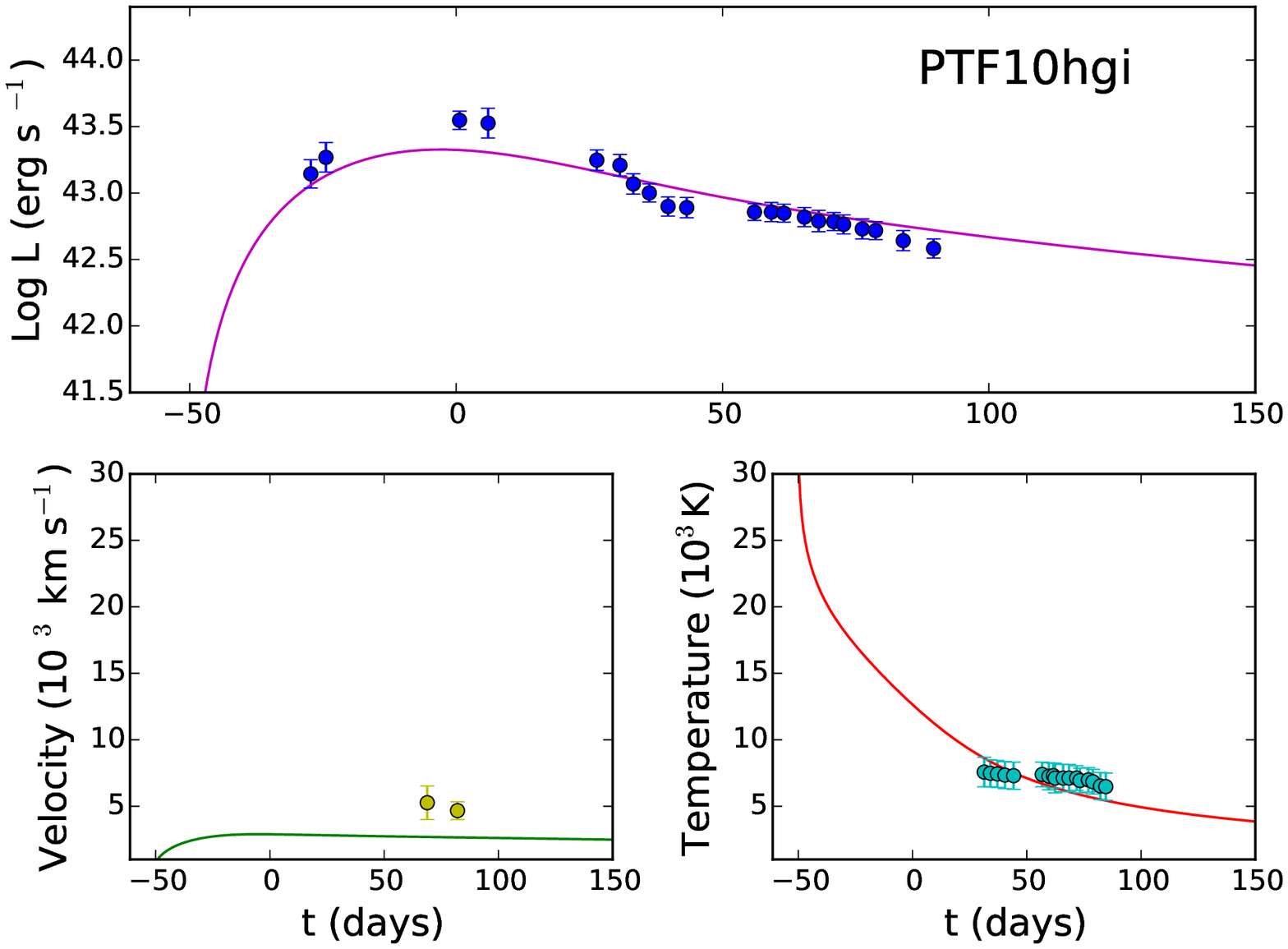}
\par
\includegraphics[width=0.4\textwidth,angle=0]{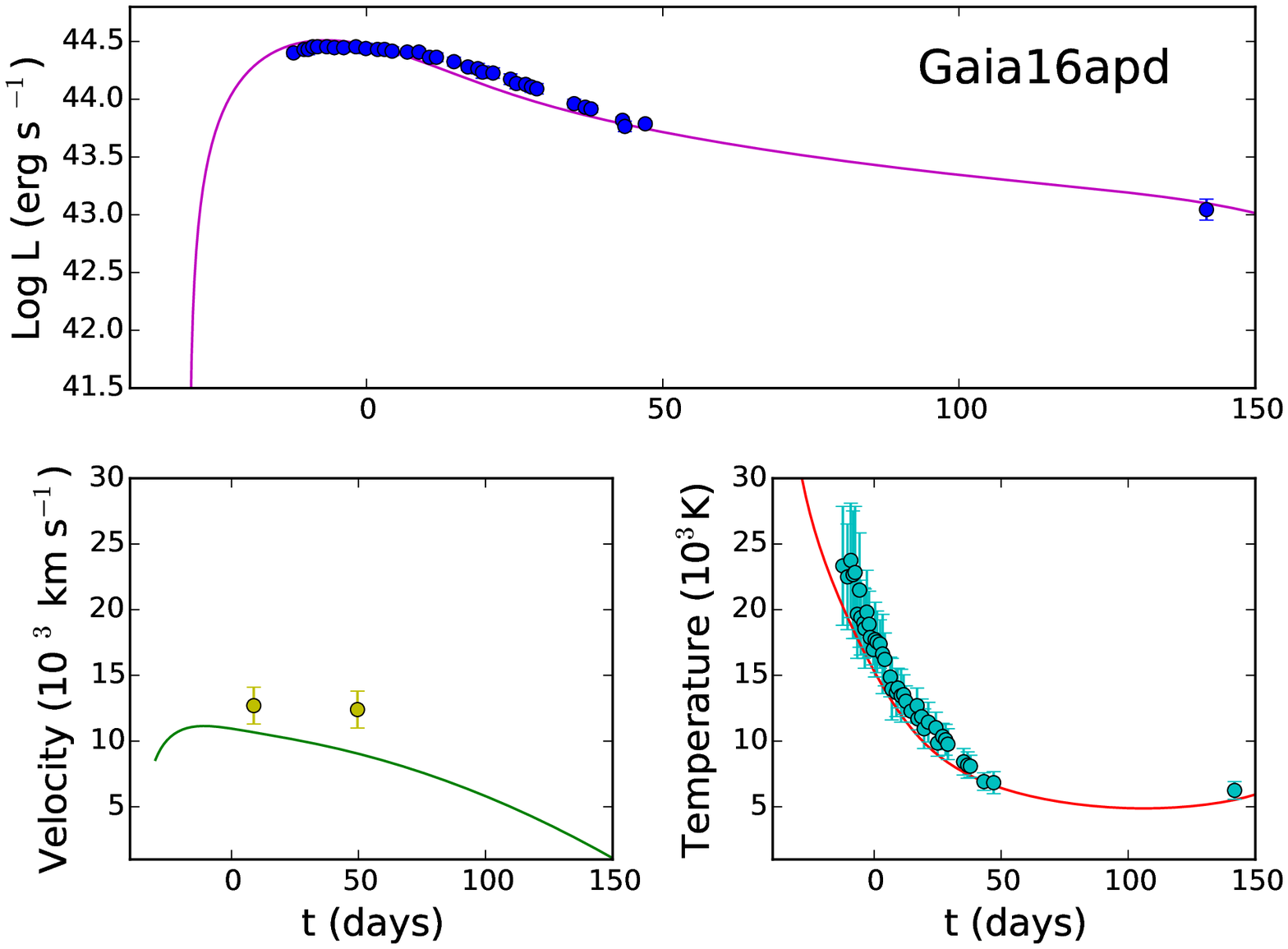}
\includegraphics[width=0.43\textwidth,angle=0]{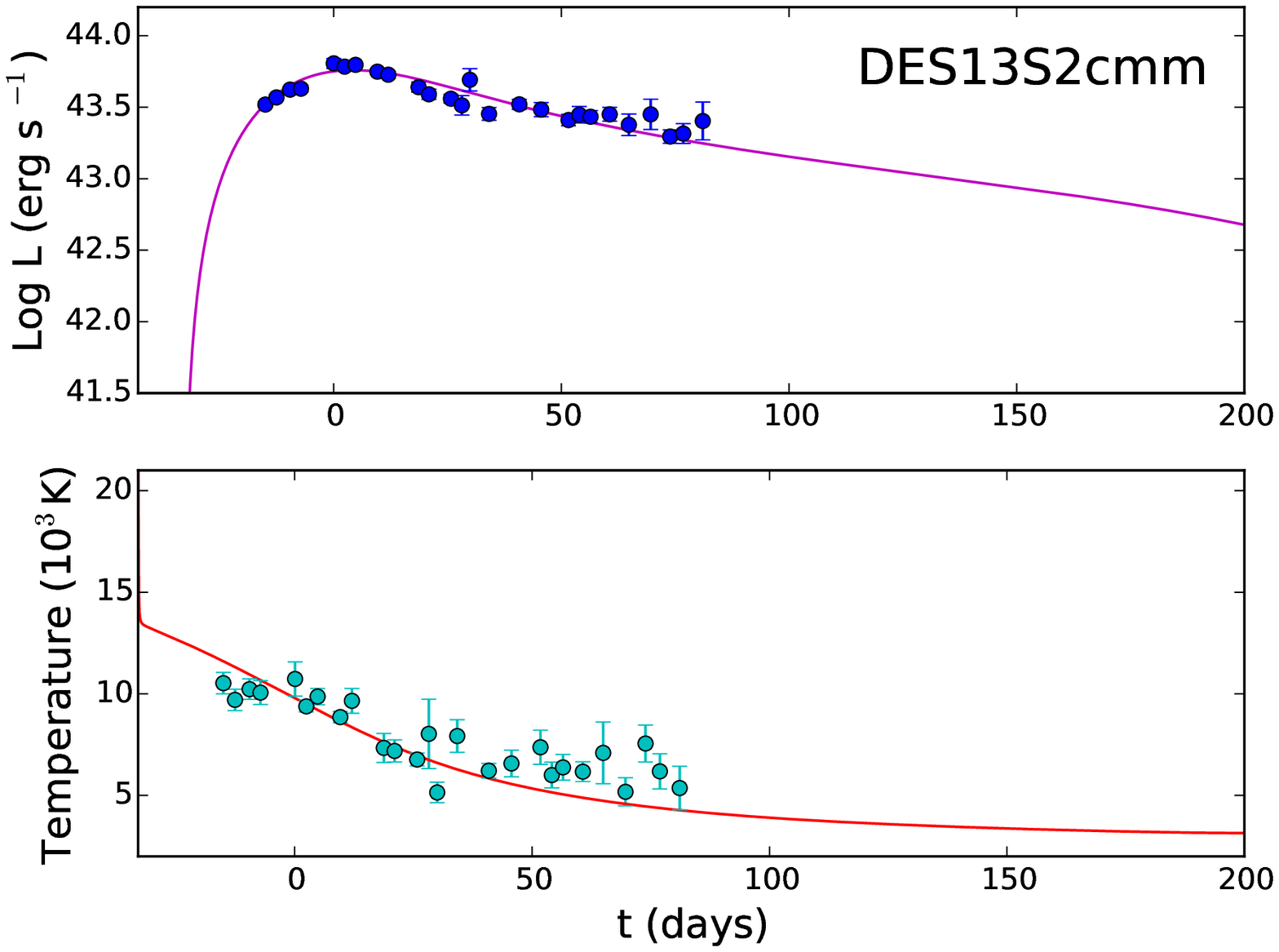}
\end{center}
\caption{Model fits for light curves, temperature evolution, and velocity
evolution of SN 2012il, PTF12dam, PTF11rks, SN 2013dg, SSS120810, PTF10hgi,
Gaia16apd, and DES13S2cmm. Parameters are shown in Table \protect\ref%
{tbl:para}.}
\label{fig:fit2}
\end{figure}

\clearpage

\begin{figure}[tbph]
\begin{center}
\includegraphics[width=0.44\textwidth,angle=0]{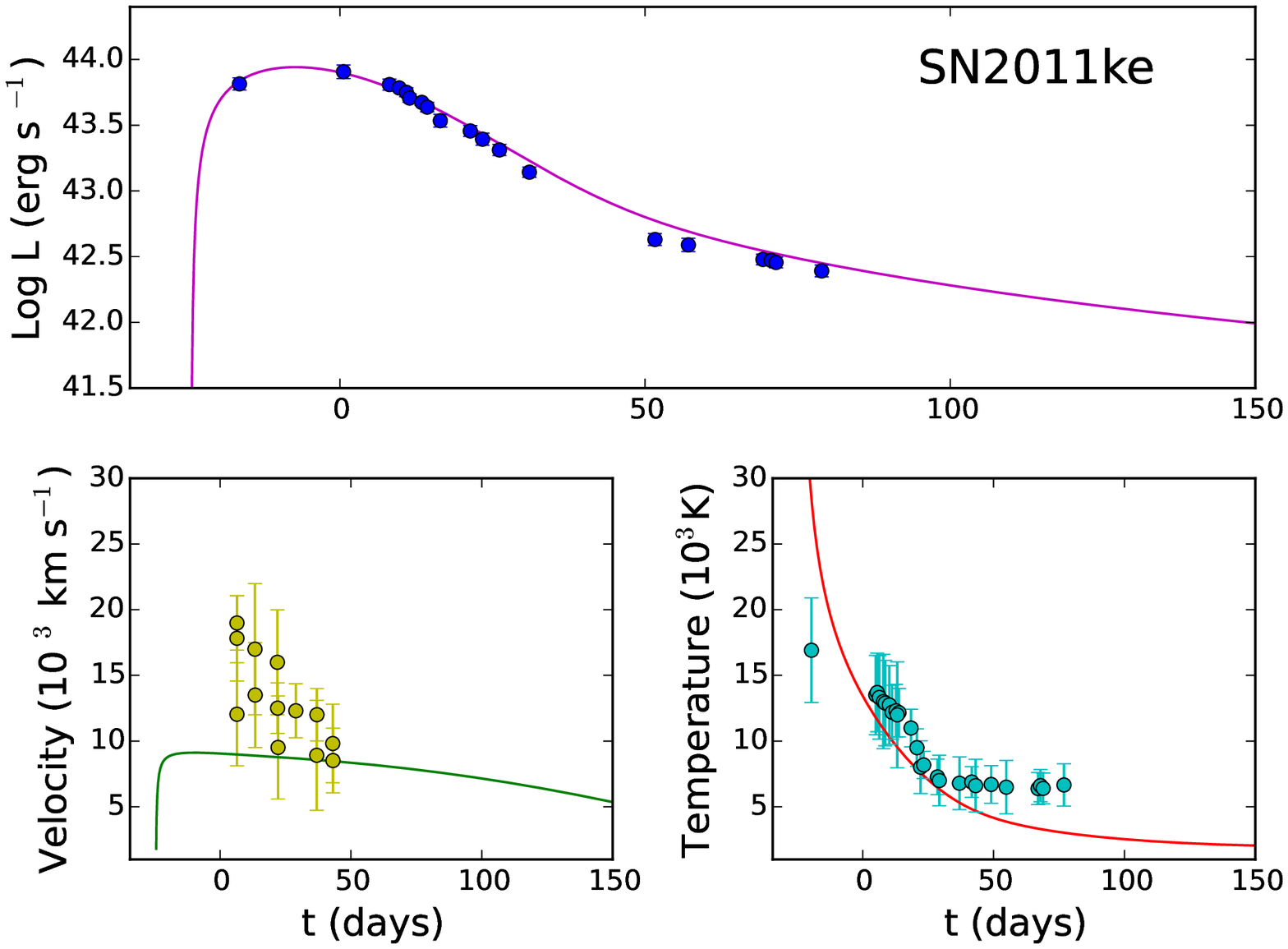}
\includegraphics[width=0.44\textwidth,angle=0]{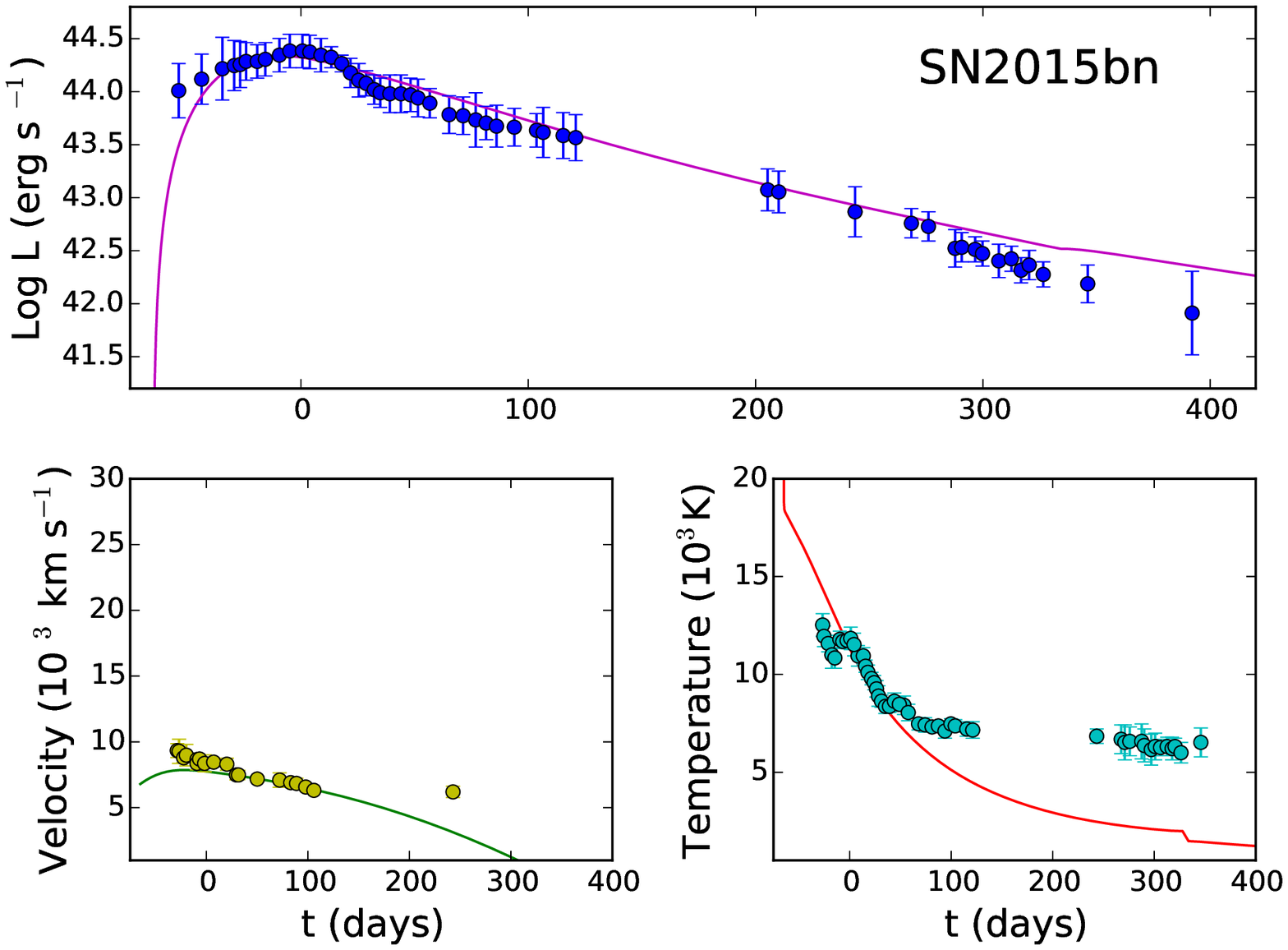}
\par
\includegraphics[width=0.44\textwidth,angle=0]{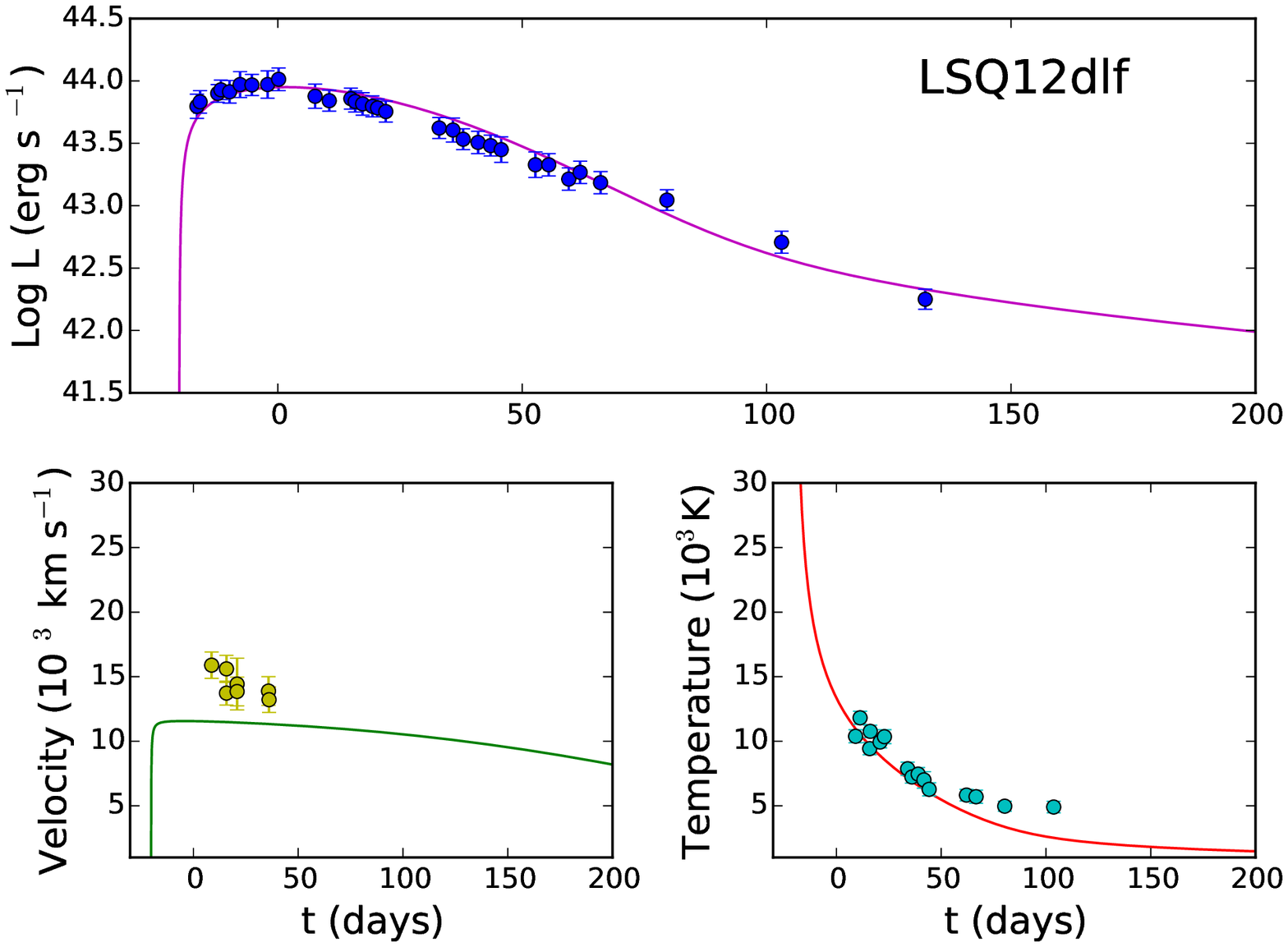}
\includegraphics[width=0.44\textwidth,angle=0]{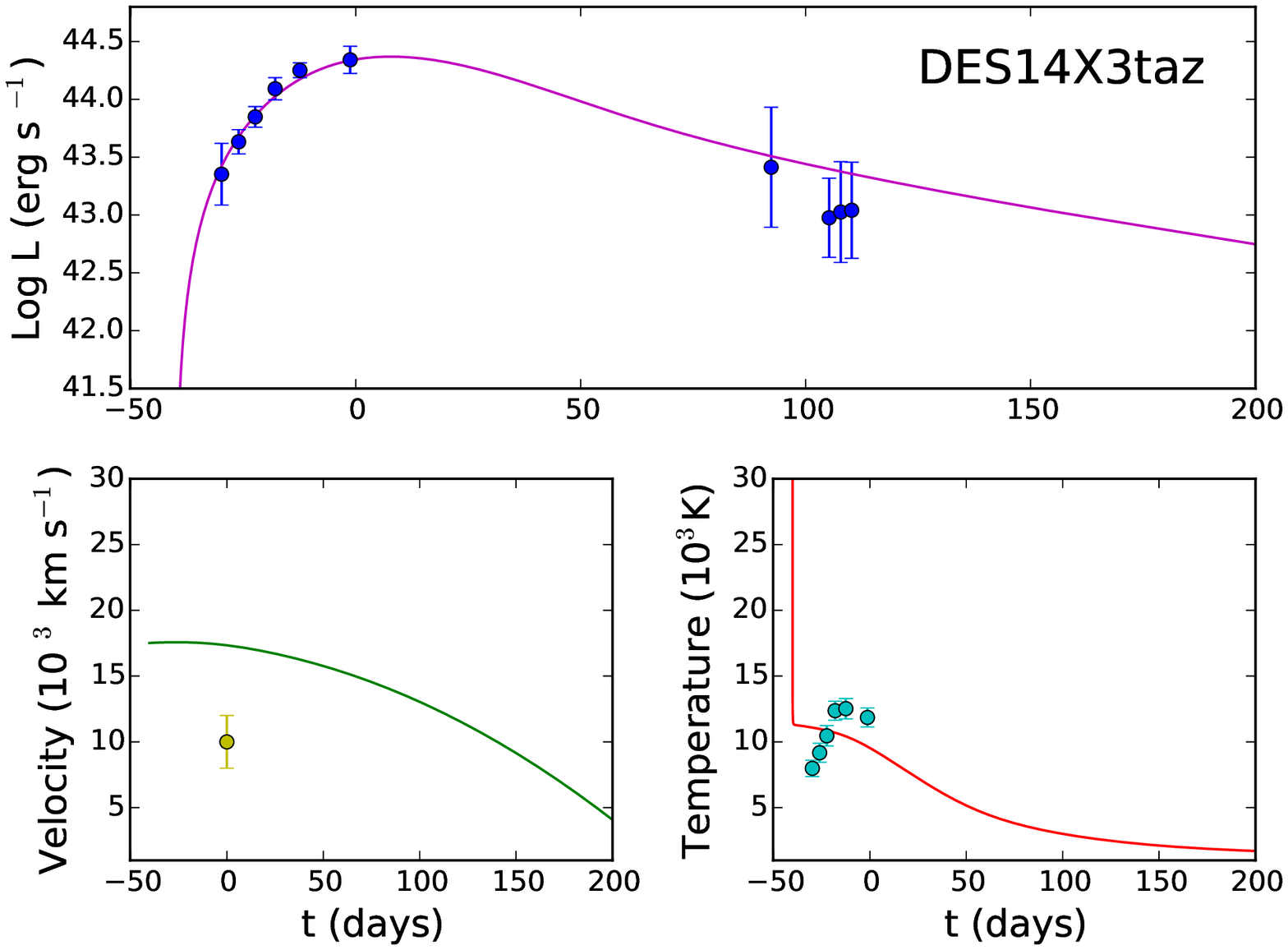}
\par
\includegraphics[width=0.44\textwidth,angle=0]{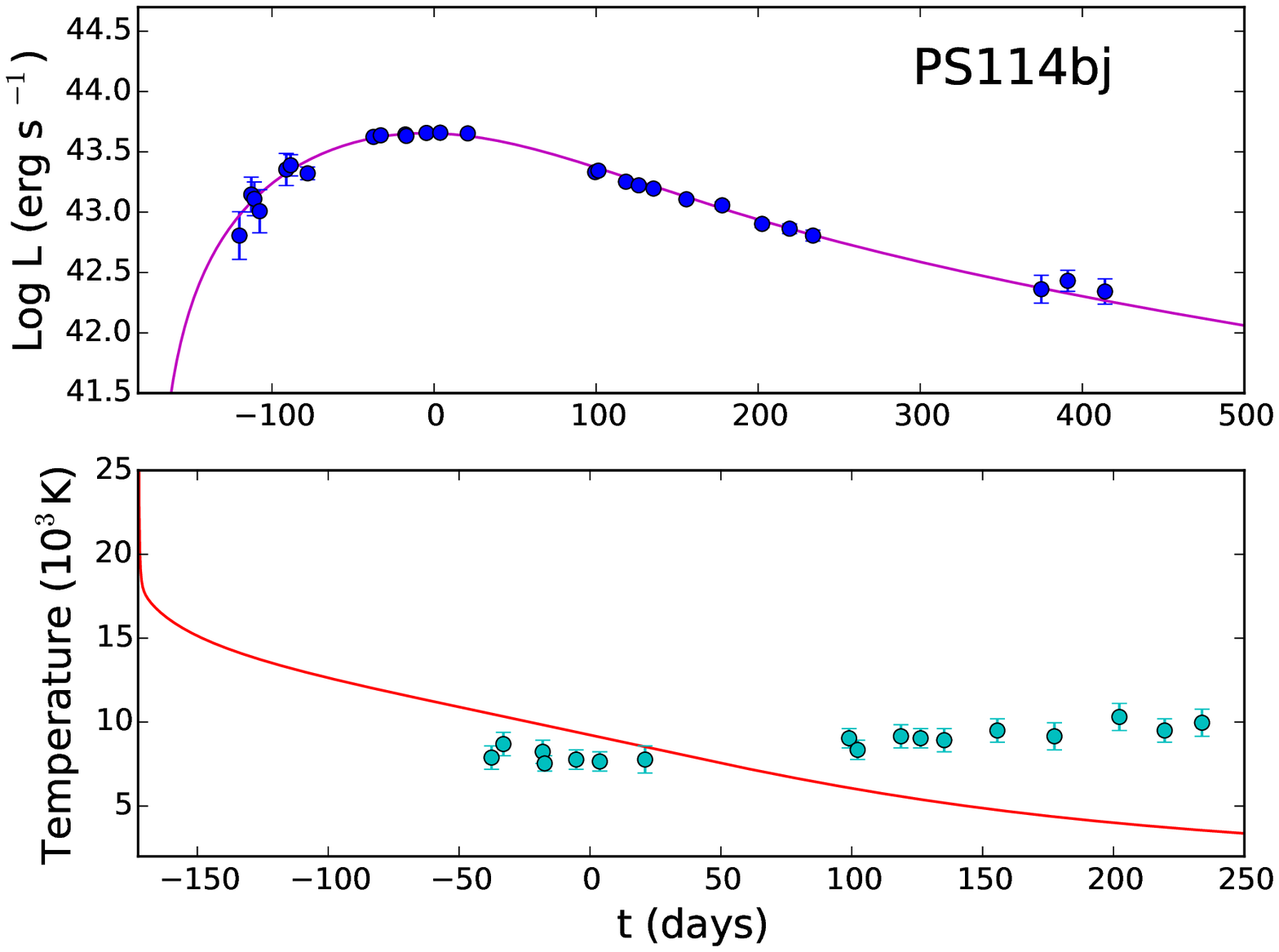}
\end{center}
\caption{Model fits for light curves, temperature evolution, and velocity
evolution of SN 2011ke, SN 2015bn, LSQ12dlf, DES14X3taz, and PS1-14bj.
Parameters are shown in Table \protect\ref{tbl:para}.}
\label{fig:fit3}
\end{figure}

\clearpage

\begin{figure}[tbph]
\begin{center}
\includegraphics[width=0.45\textwidth,angle=0]{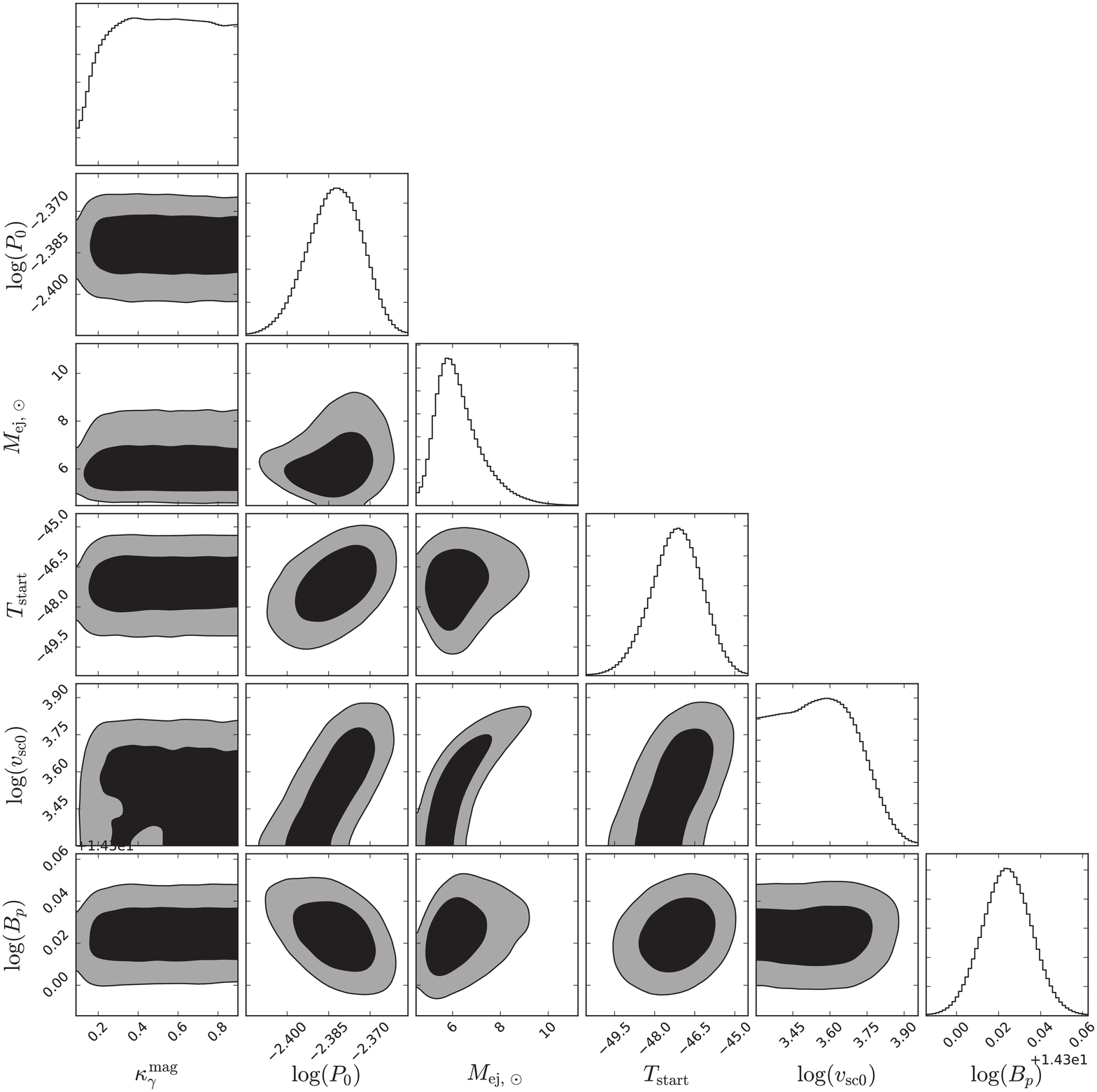}
\includegraphics[width=0.45\textwidth,angle=0]{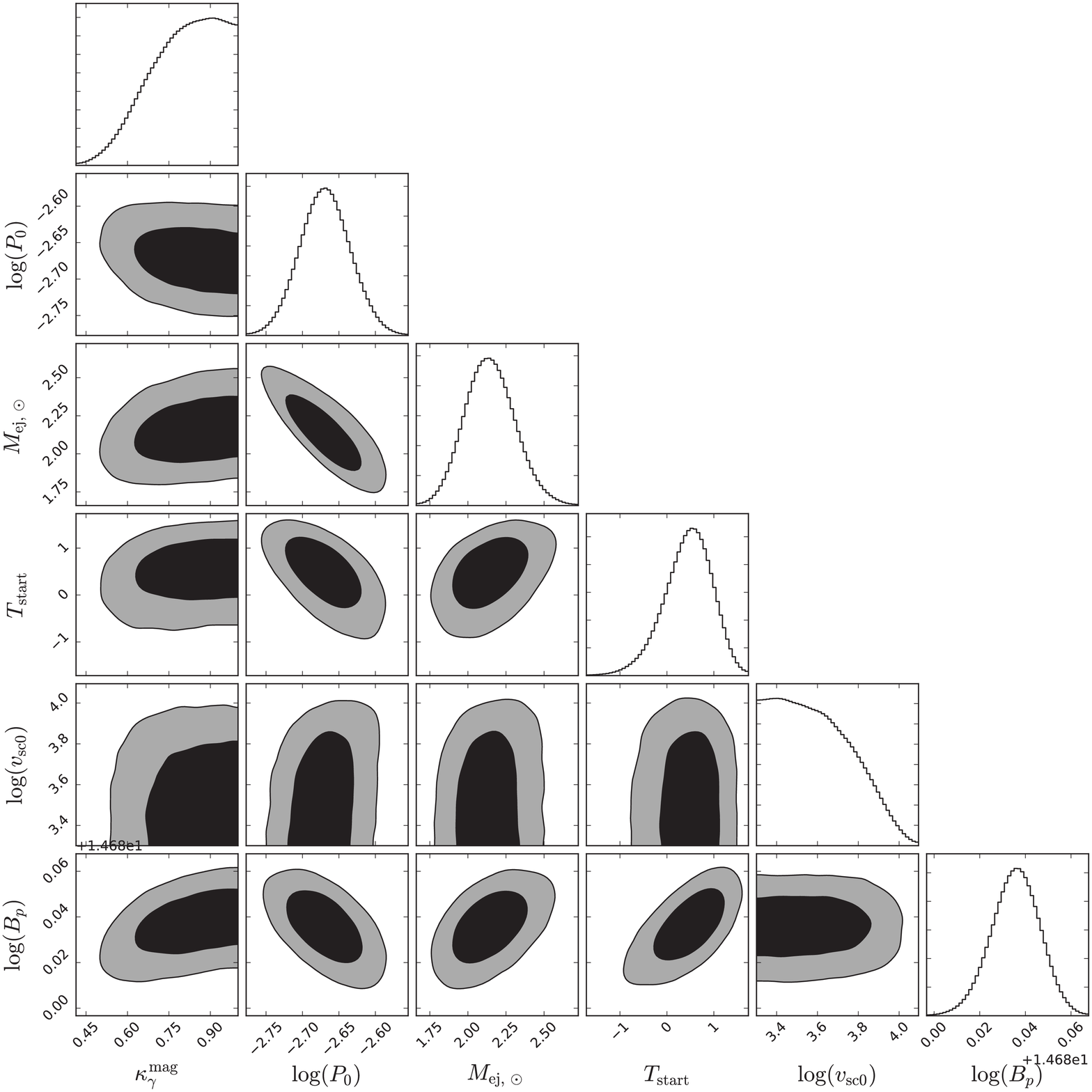}
\par
\includegraphics[width=0.45\textwidth,angle=0]{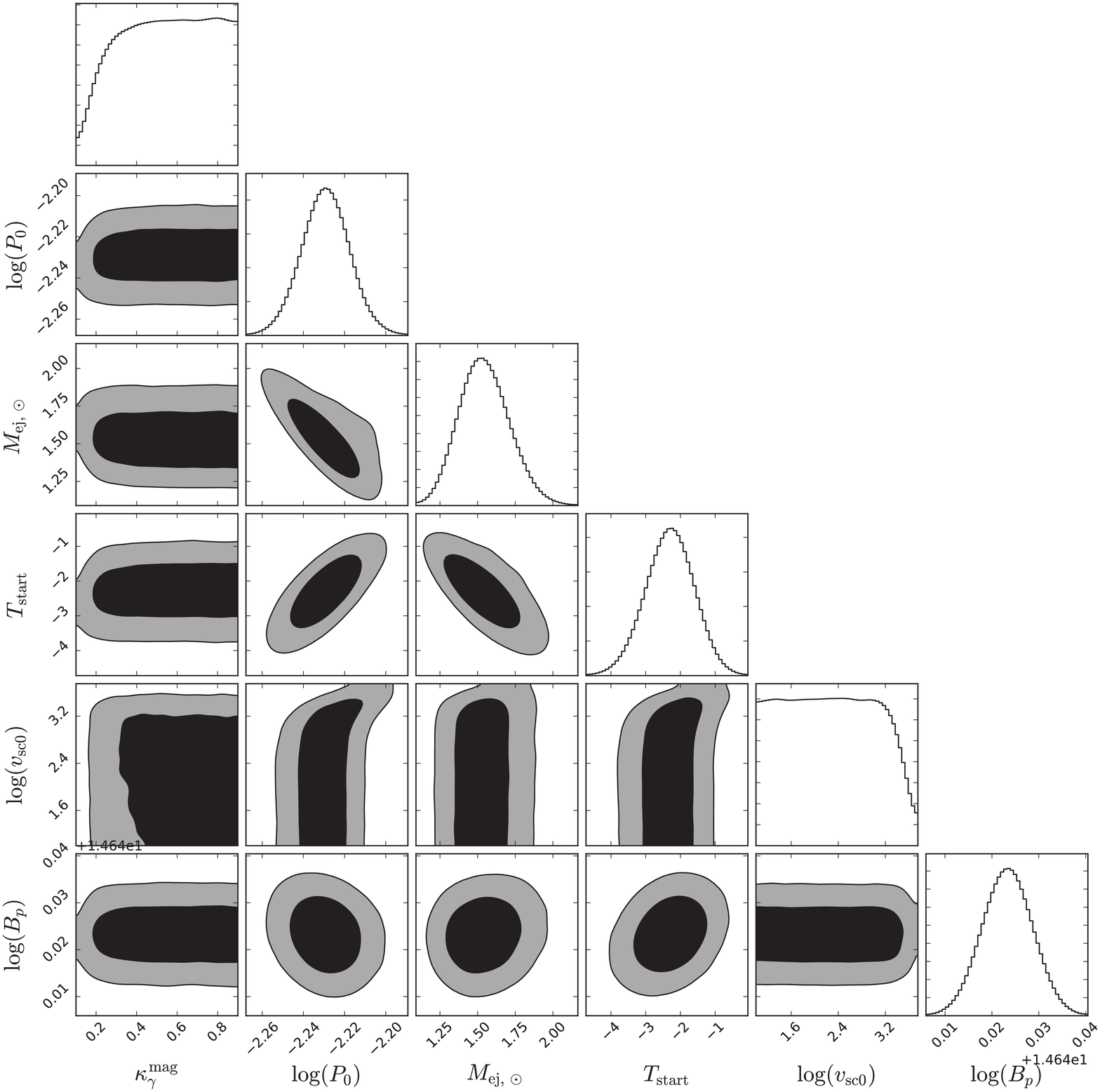}
\includegraphics[width=0.45\textwidth,angle=0]{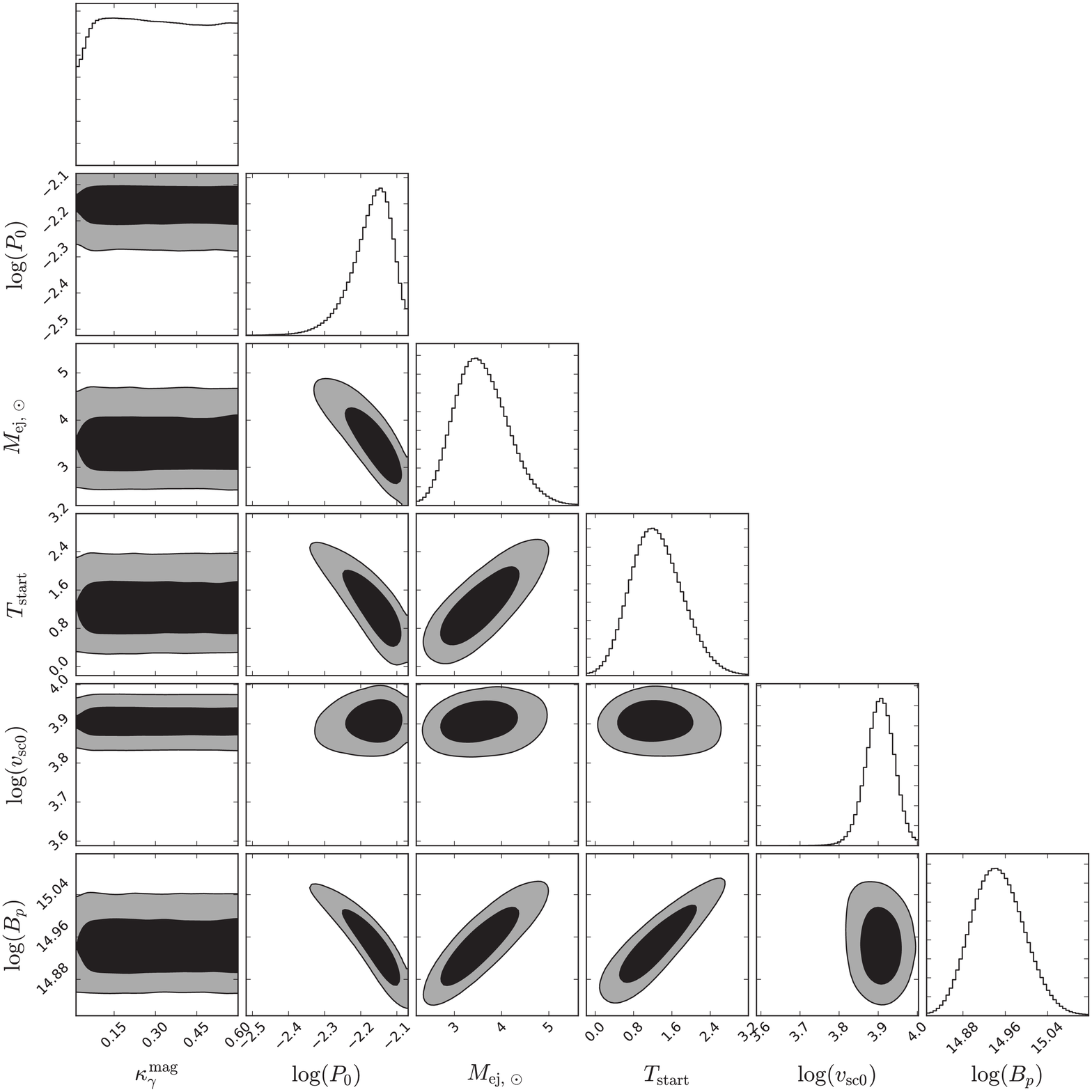}
\end{center}
\caption{Confidence contours of parameter corner in the modeling of
PS1-11ap, SN 2011kf, SN2012il and PTF11rks. The contours are 1 $\protect%
\sigma $ and 2 $\protect\sigma $ uncertainties, respectively.}
\label{fig:corner}
\end{figure}
\clearpage
\begin{figure}[tbph]
\begin{center}
\includegraphics[width=0.4\textwidth,angle=0]{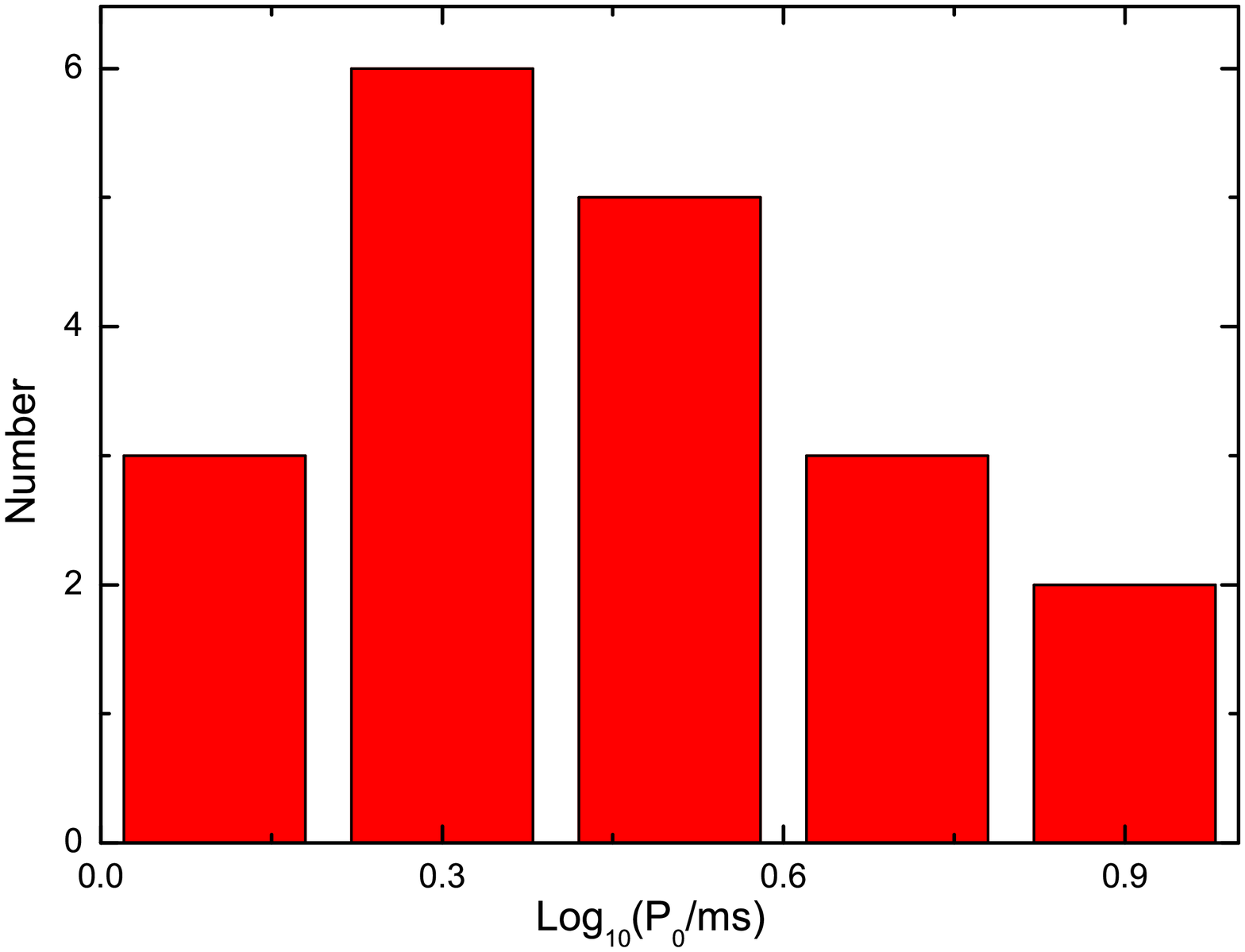}
\includegraphics[width=0.4\textwidth,angle=0]{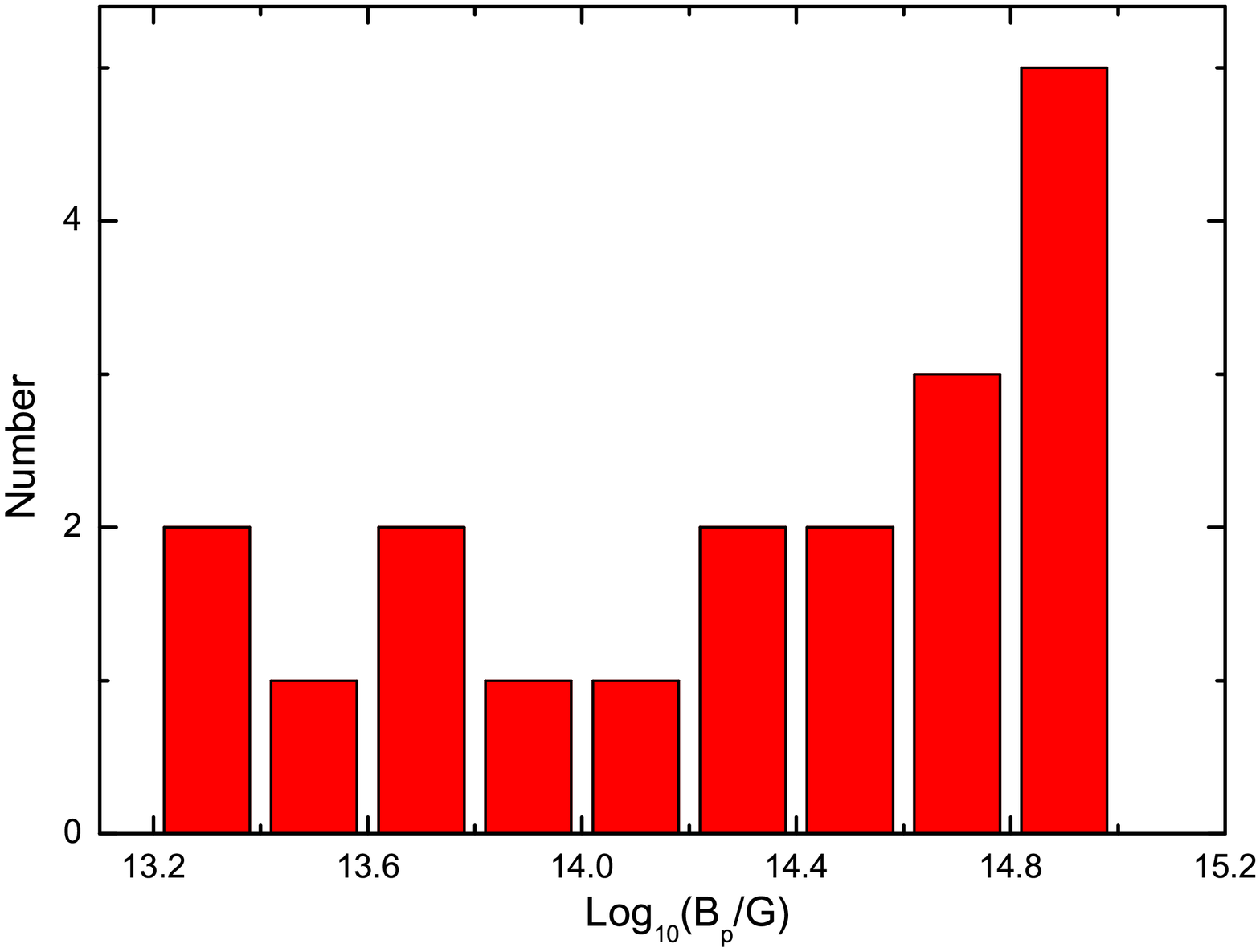}
\par
\includegraphics[width=0.4\textwidth,angle=0]{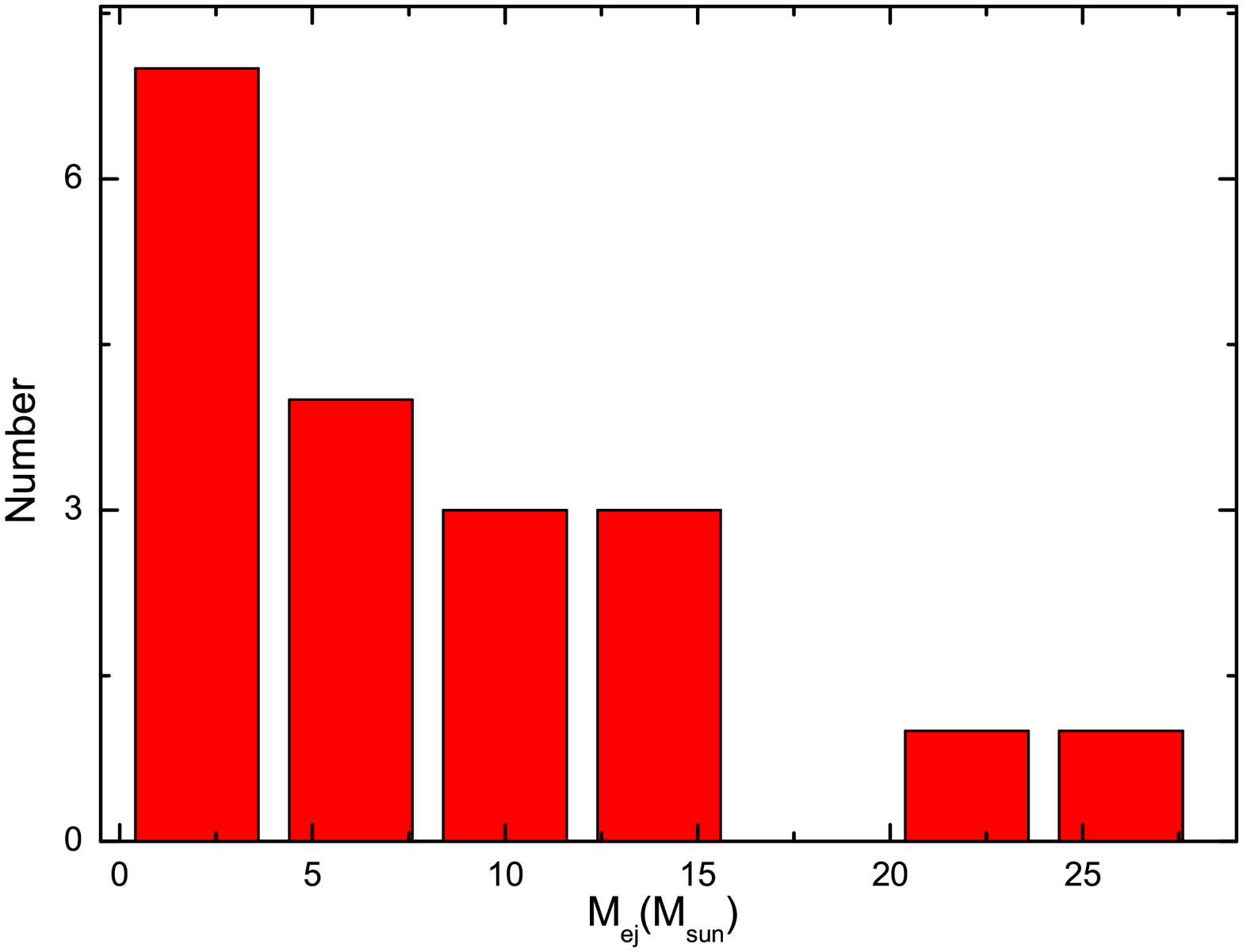}
\includegraphics[width=0.4\textwidth,angle=0]{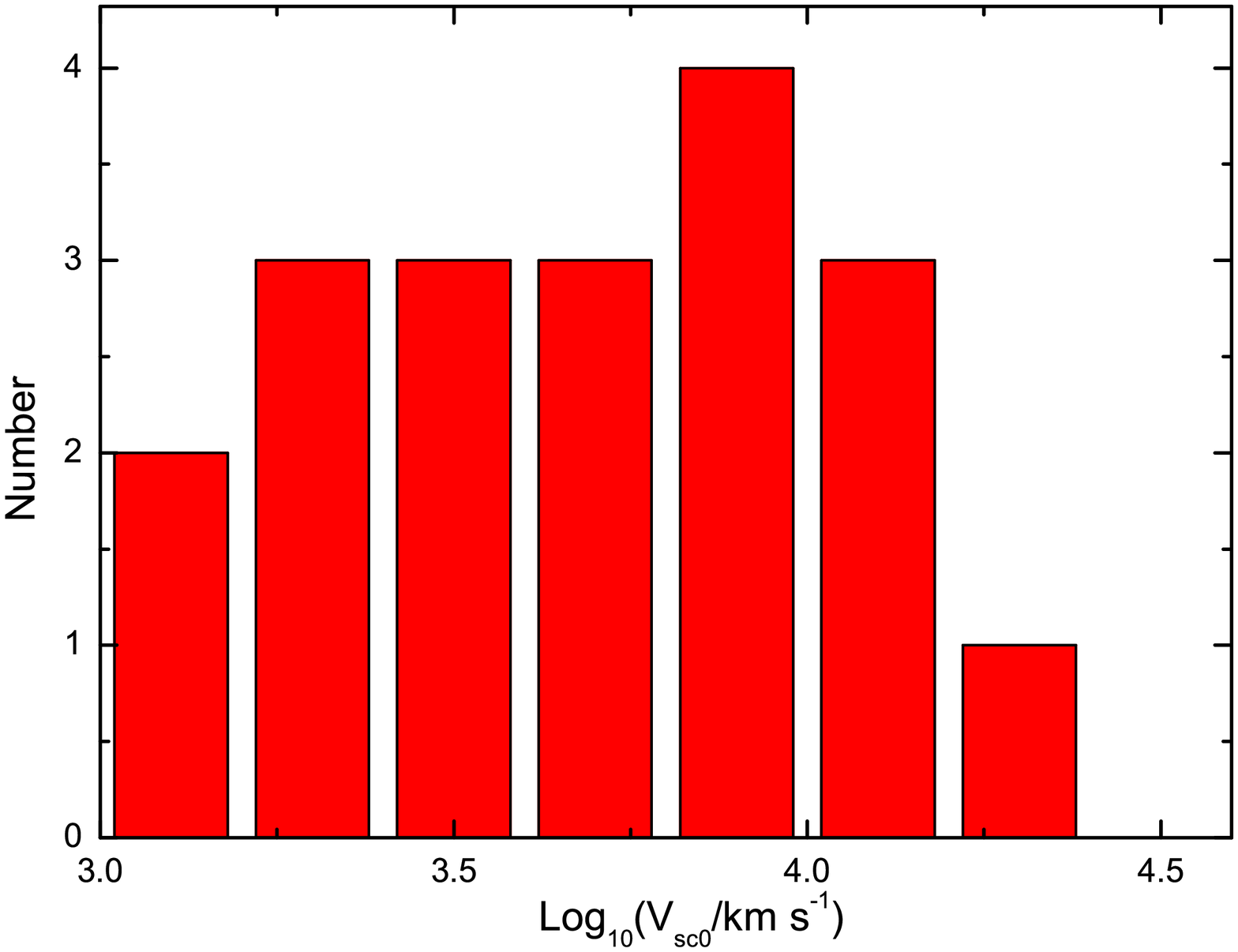}
\par
\includegraphics[width=0.4\textwidth,angle=0]{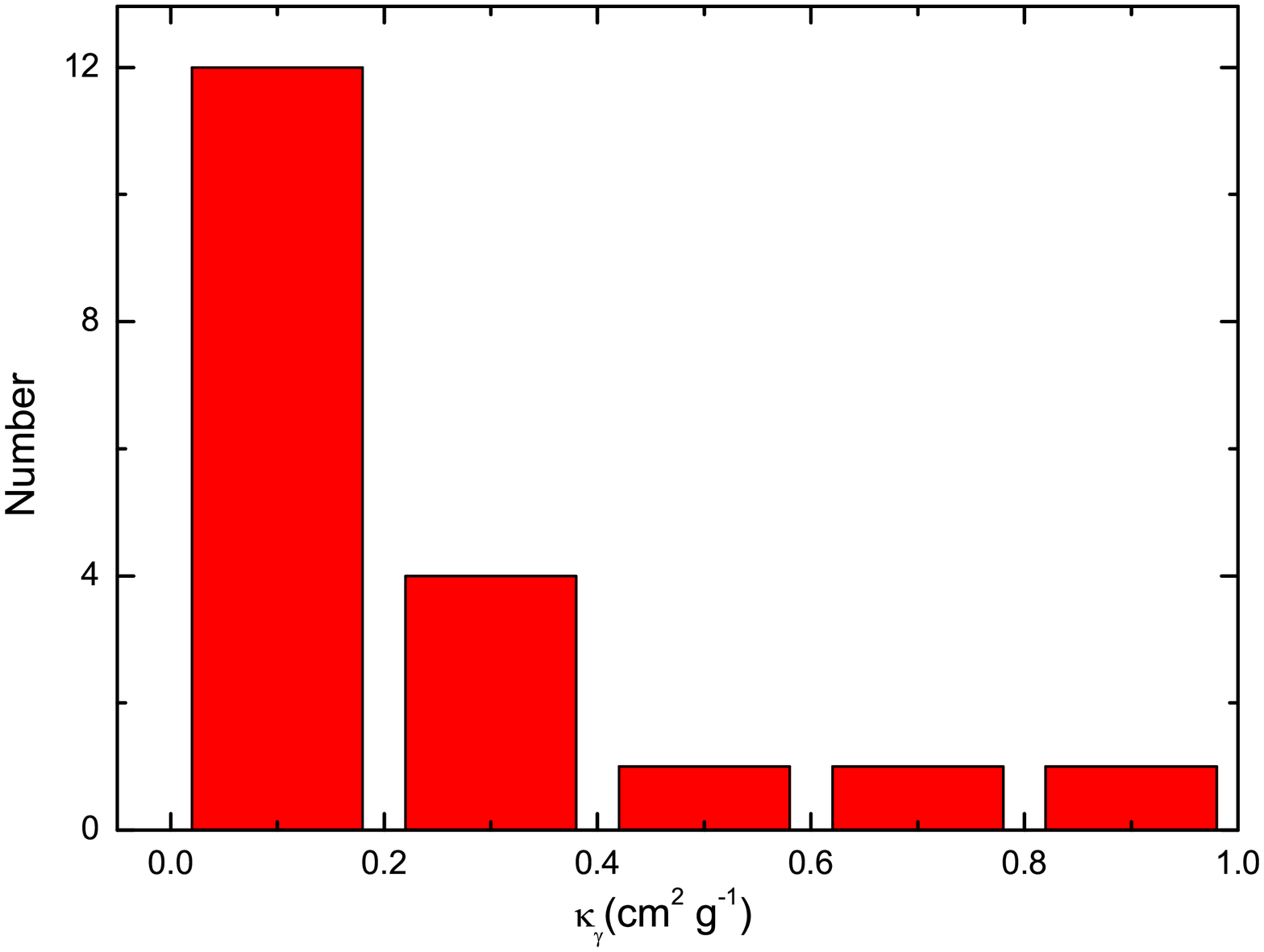}
\includegraphics[width=0.4\textwidth,angle=0]{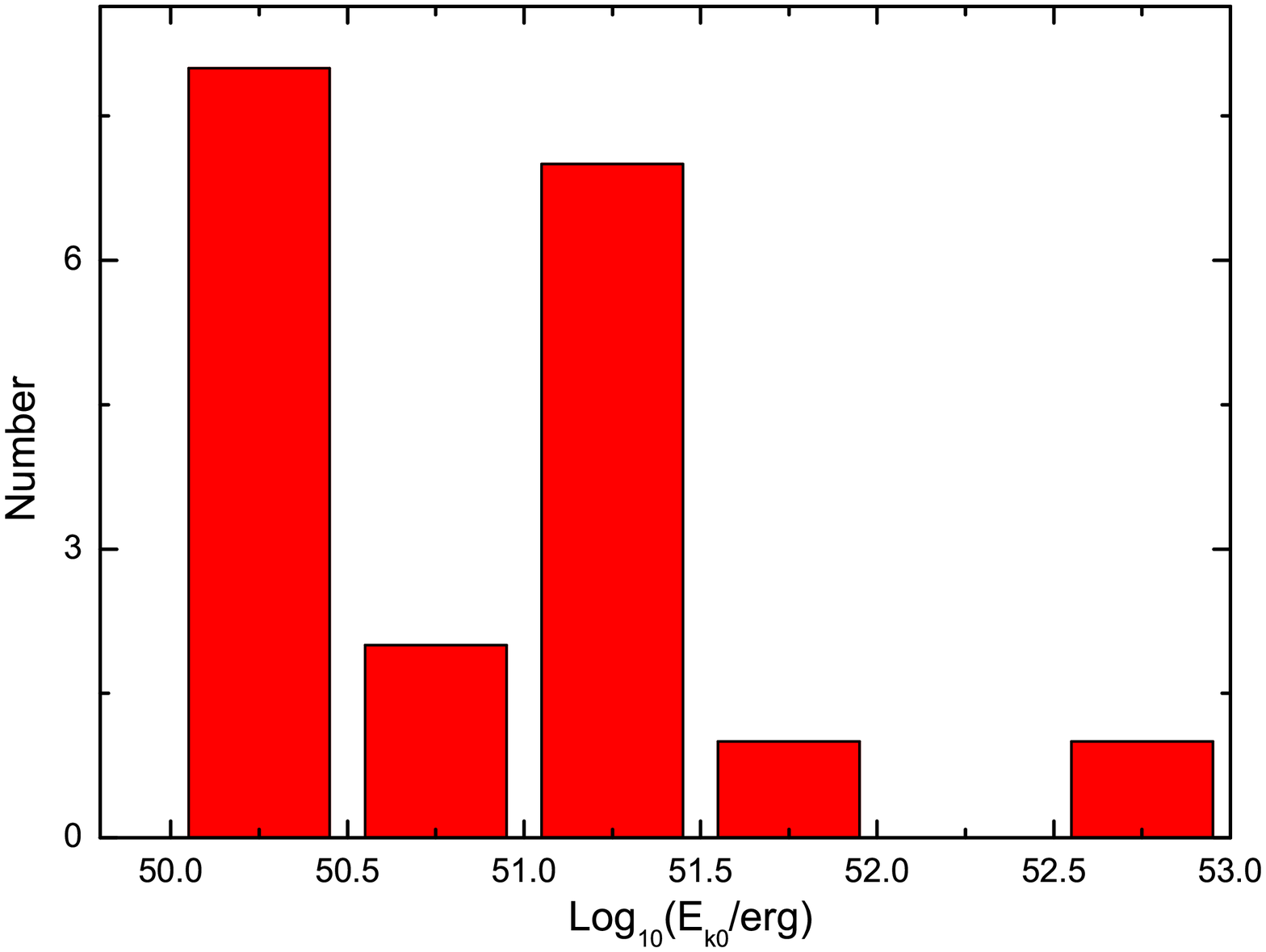}
\par
\includegraphics[width=0.4\textwidth,angle=0]{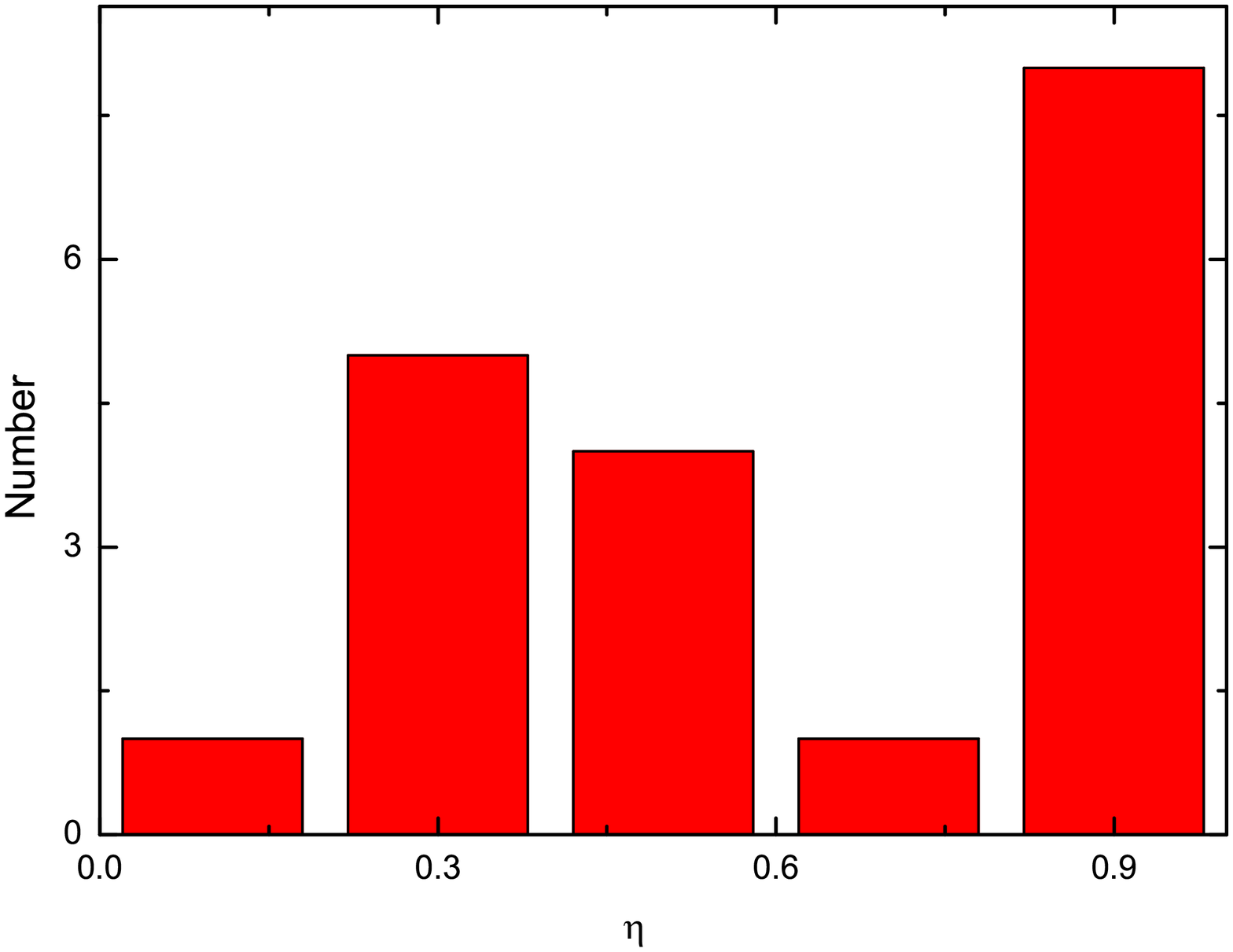}
\end{center}
\caption{Distributions of the initial stellar periods $P_{0}$, magnetic
strength $B_{p}$, ejecta masses $M_{\text{ej}}$, initial scale velocities of
the ejecta $v_{\text{sc0}}$, the gamma-ray opacity $\protect\kappa _{\protect%
\gamma }$, initial kinetic energy of the ejecta $E_{\text{K0}}$ as well as
the accumulative fraction of the rotational energy of the magnetar converted
to the kinetic energy $\protect\eta $ in the modeling of the sample.}
\label{fig:dis}
\end{figure}

\clearpage
\begin{figure}[tbph]
\begin{center}
\includegraphics[width=0.45\textwidth,angle=0]{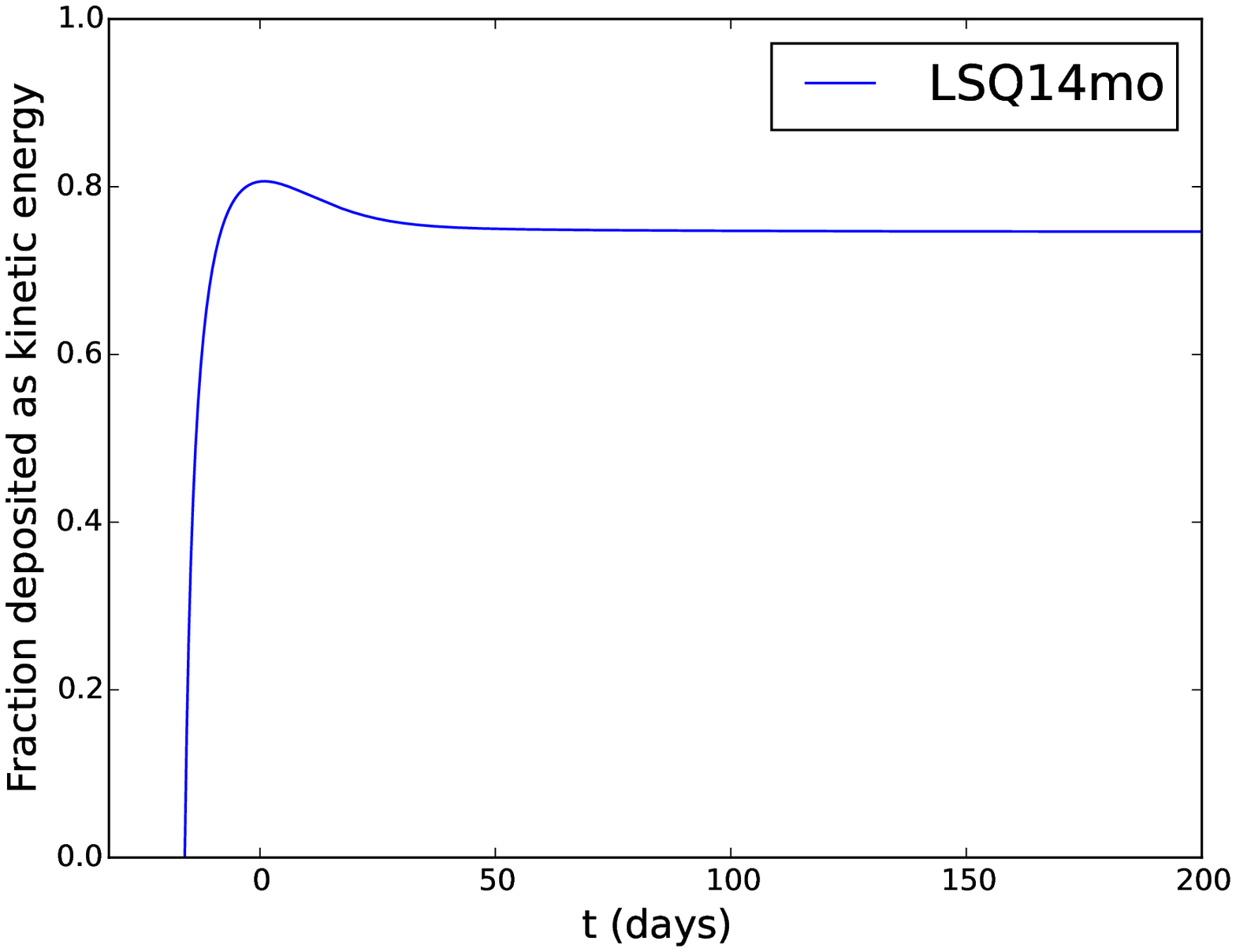}
\includegraphics[width=0.45\textwidth,angle=0]{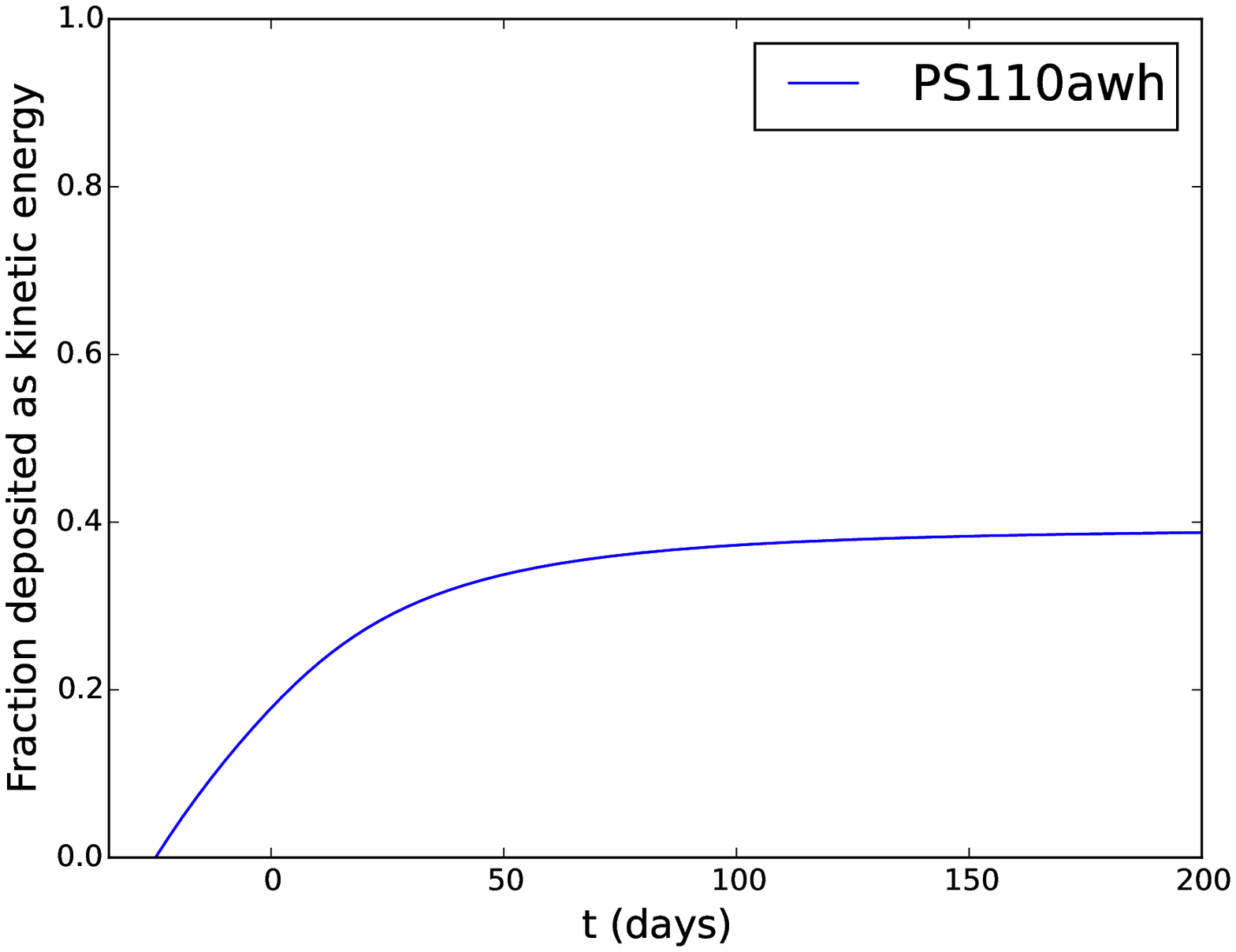}
\par
\includegraphics[width=0.45\textwidth,angle=0]{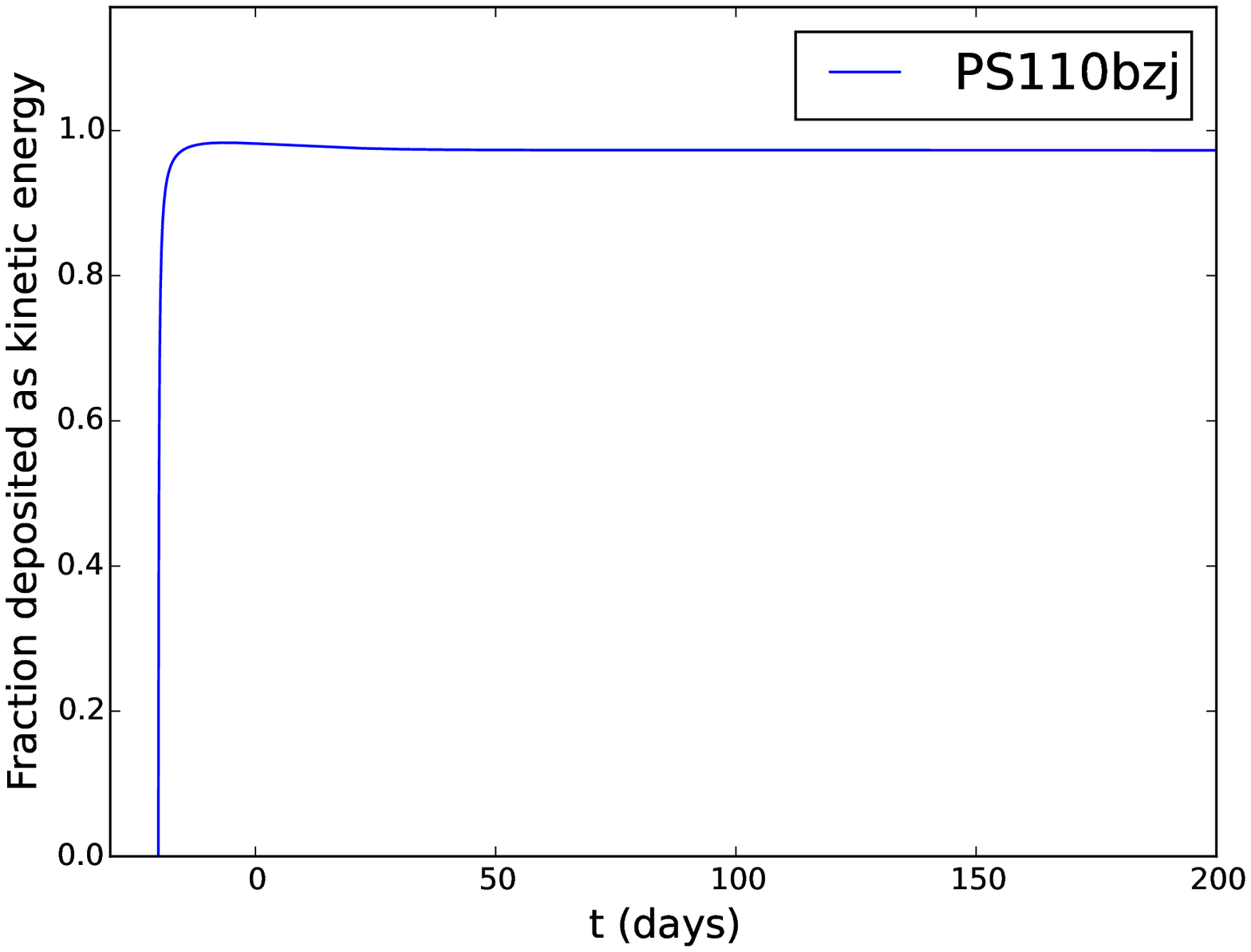}
\includegraphics[width=0.45\textwidth,angle=0]{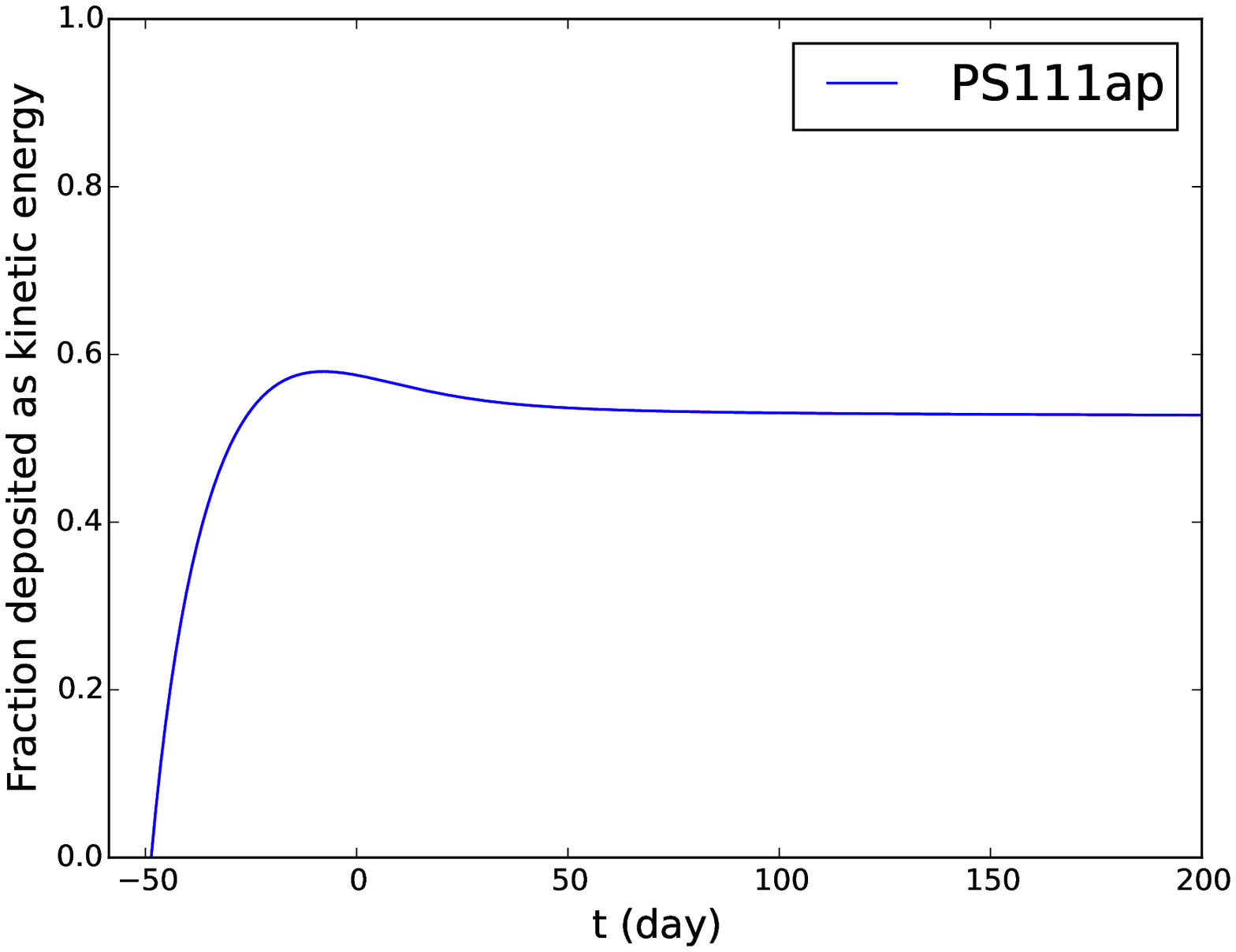}
\par
\includegraphics[width=0.45\textwidth,angle=0]{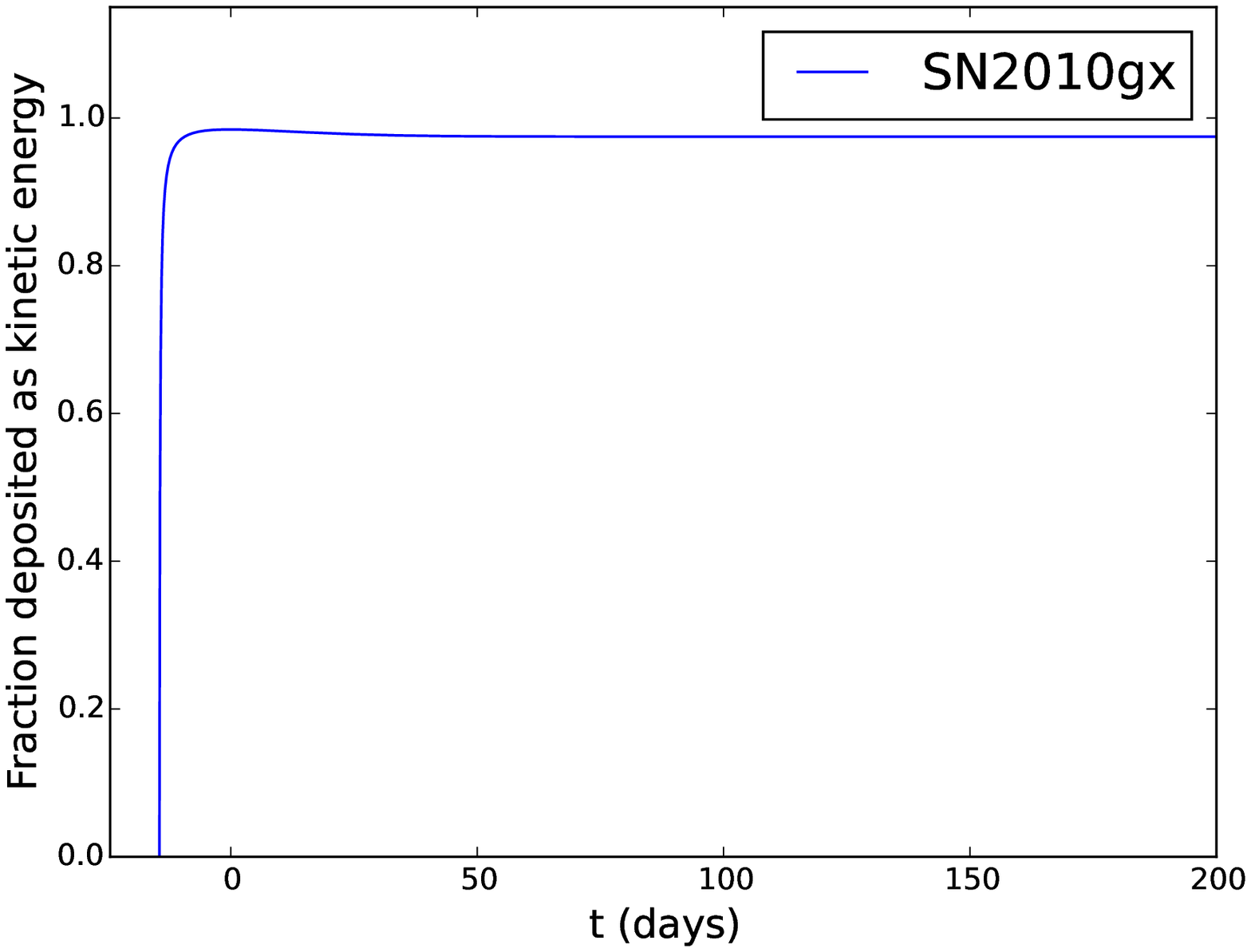}
\includegraphics[width=0.45\textwidth,angle=0]{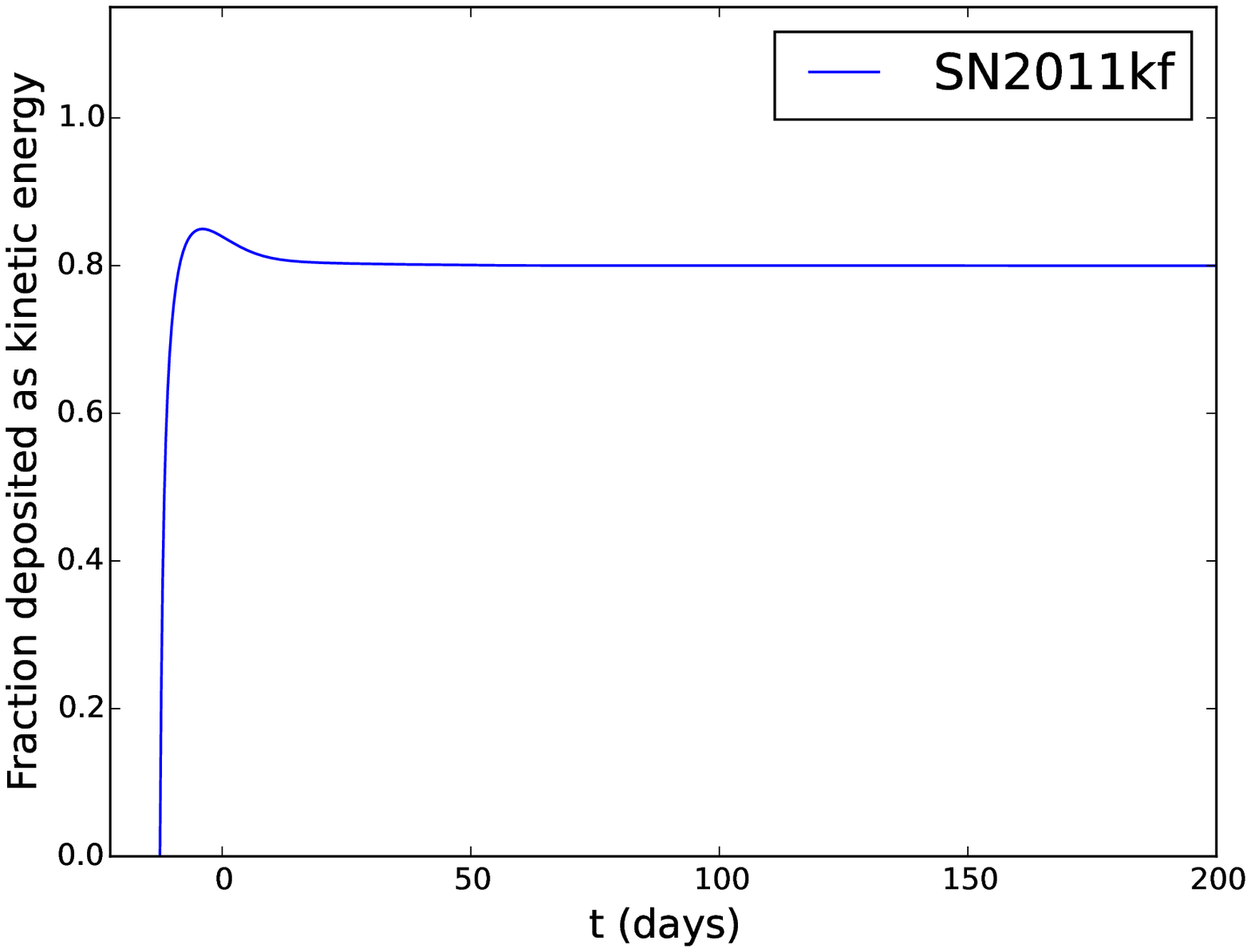}
\end{center}
\caption{The accumulative fraction of the stellar rotational energy
converted to the kinetic energy of LSQ14mo, PS1-10awh, PS1-10bzj, PS1-11ap,
SN 2010gx, and SN 2011kf.}
\label{fig:fra}
\end{figure}
\clearpage
\begin{figure}[tbph]
\begin{center}
\includegraphics[width=0.6\textwidth,angle=0]{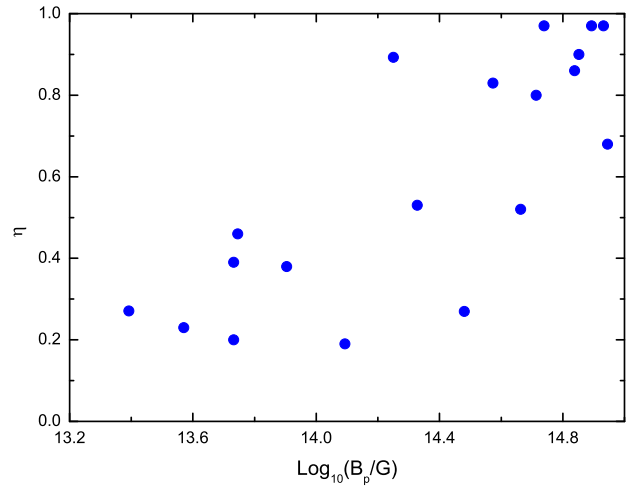}
\end{center}
\caption{The magnetic strength ($B_{p}$) versus the final fraction ($\protect%
\eta $) of the magnetar's rotational energy converted to the kinetic energy.
Parameters are listed in Table \protect\ref{tbl:derived parameters}.}
\label{fig:etavsb}
\end{figure}

\end{document}